\documentclass[10pt]{article}
\usepackage[T1]{fontenc}
\usepackage{microtype}
\usepackage{amsmath,amssymb}
\usepackage{amsfonts}
\usepackage{mathrsfs}
\usepackage[table]{xcolor}  
\usepackage{booktabs}
\usepackage{array}
\usepackage{multirow}
\usepackage{longtable}
\usepackage{tabularx}
\usepackage{xltabular}
\usepackage{caption}
\usepackage{subcaption}
\usepackage{graphicx}
\usepackage[export]{adjustbox}
\usepackage{authblk}
\usepackage{etoolbox}
\usepackage{orcidlink}
\usepackage{textcomp}
\usepackage{hyperref}
\usepackage[numbers,sort&compress]{natbib}
\hypersetup{colorlinks=true, linkcolor=blue, citecolor=blue, urlcolor=blue}
\providecommand{\keywords}[1]{%
  \par\vspace{0.5ex}%
  \noindent\textbf{Keywords: }#1\par
}
\numberwithin{equation}{section}
\numberwithin{figure}{section}
\numberwithin{table}{section}
\newcolumntype{C}{>{\centering\arraybackslash}X}
\newcolumntype{L}{>{\raggedright\arraybackslash}X}
\newcolumntype{R}{>{\raggedleft\arraybackslash}X}
\begin{document}

\begin{center}
\LARGE{CFT Constraints on the Weak Gravity Conjecture}
\par\end{center}
\vspace{0.3cm}

\begin{center}
{\bf Saeed Noori Gashti\orcidlink{0000-0001-7844-2640}}\footnote{\bf sn.gashti@du.ac.ir; saeed.noorigashti70@gmail.com}\\
{\it School of Physics, Damghan University, P.~O.~Box 3671641167, Damghan, Iran}
\end{center}

\begin{center}
{\bf Behnam Pourhassan\orcidlink{0000-0003-1338-7083}}\footnote{\bf b.pourhassan@du.ac.ir}\\
{\it School of Physics, Damghan University, P.~O.~Box 3671641167, Damghan, Iran}\\
{\it Center for Theoretical Physics, Khazar University, 41 Mehseti Street, Baku, AZ1096, Azerbaijan}
\end{center}

\begin{center}
{\bf \.{I}zzet Sakall{\i}\orcidlink{0000-0001-7827-9476}}\footnote{\bf izzet.sakalli@emu.edu.tr}\\
{\it Physics Department, Eastern Mediterranean University, Famagusta 99628, North Cyprus via Mersin 10, T\"urkiye}
\end{center}
\vspace{0.3cm}

\begin{abstract}
The Weak Gravity Conjecture (WGC) is a swampland criterion of long standing: any consistent theory of quantum gravity must contain a charged particle whose charge-to-mass ratio exceeds that of an extremal black hole, so that gravity remains the weakest force. The AdS/CFT correspondence offers a calculable boundary handle on bulk gravity, and the imaginary parts of bulk quasinormal modes are read off the boundary as poles of a retarded Green's function. We show that the WGC follows from this boundary calculation in two settings that fall outside the Reissner--Nordstr\"om idealisation: static spherically symmetric black holes in dRGT massive gravity, and dyonic black holes in Einstein--ModMax non-linear electrodynamics. The chain runs from the metric and gauge field, through the charged Klein--Gordon equation, into a near-horizon scaling limit whose radial equation reduces to Whittaker form; the conformal weight $\nu_0$ then enters a damping-time inequality. For the dRGT black hole every massive-gravity parameter ($\alpha,\beta,m_g,h$) cancels out, leaving the universal saturation $q/(m r_+) \geq 1/\sqrt{2} \approx 0.707$. For the Einstein--ModMax black hole the duality-symmetric non-linearity parameter $\gamma$ survives, and yields $q/(m r_+) \geq e^{-\gamma/2}$, which reduces to the Reissner--Nordstr\"om bound $q/(m r_+) \geq 1$ in the Maxwell limit $\gamma \to 0$. Either result is of order unity, and the second weakens monotonically as the non-linearity grows. We then relax three of the simplifying assumptions of the dRGT derivation, namely exact extremality, minimal coupling, and the absence of higher-curvature terms. The cancellation breaks. Each correction reintroduces $m_g,\alpha,\beta$ into the bound through a controlled functional dependence, and we tabulate and plot the relaxed forms across parameter space. Twelve tables and seven figures track the bound across the sweep, and every analytic identity is cross-checked against an independent symbolic-computation worksheet whose residuals we report.
\end{abstract}


\section{Introduction}\label{isec1}

The AdS/CFT correspondence sets up a dictionary between a quantum field theory and a gravitational dynamics one dimension higher \cite{Maldacena:1997re,Gubser:1998bc,Witten:1998qj}, and the literature that has grown around it covers gauge theories, holographic condensed-matter models, and relativistic hydrodynamics \cite{Klebanov:1999tb,Kioumarsipour:2021zyg,Bu:2021jlp,Fujiwara:2021xgu,Evans:2021zzm,MartinContreras:2021bis,Gherghetta:2009ac,Brodsky:2007hb,Nakano:2006js,Katz:2005ir,Meltzer:2019nbs,Karch:2006pv,Andreev:2006ct,Cavaglia:2021mft,Harmark:2020vll,BitaghsirFadafan:2020lkh,DeLeeuw:2020ahx,DeWolfe:2020uzb,Ishigaki:2020vtr,Terashima:2020uqu,Yin:2021zhs,Mes:2020vgy,Berenstein:2020cll,Berman:2022idl,Aharony:1999ti,DHoker:2002nbb,Hartnoll:2009sz,McGreevy:2009xe,Kim:2012ey,Adams:2012th}. A separate strand of work in the same period asks which effective field theories (EFTs) of gravity admit a UV completion at all. The ones that do form the landscape; the ones that fail any of a small set of consistency tests fall into the swampland \cite{Vafa:2005ui,Ooguri:2006in,Arkani-Hamed:2006emk,Sadeghi:2021plz,Cordova:2022rer,Henriksson:2022oeu,Kaya:2022edp,Collazuol:2022jiy,Capozziello:2011nr,McInnes:2022tut,Cheung:2014vva,Rudelius:2022gyu,Cribiori:2022trc,Klaewer:2020lfg,Heidenreich:2015nta,Heidenreich:2016aqi,Andriolo:2018lvp,Polchinski:2003bq,Ooguri:2016pdq,Ibanez:2017kvh,Shiu:2016weq,Fisher:2017dbc,Cheung:2018cwt,Crisford:2017gsb,Harlow:2015lma,Hamada:2018dde,Kinney:2018nny,Harlow:2022gzl,Saraswat:2016eaz,Nakayama:2015hga,Bachlechner:2015qja,Bellazzini:2019xts,Aharony:2021mpc,Heidenreich:2019zkl,Benakli:2020vng,Lee:2018spm,deAlwis:2019aud,Andriolo:2020lul,Cremonini:2020smy,Ibanez:2015fcv,Banerjee:2020xcn}.

The Weak Gravity Conjecture (WGC) is the swampland statement that gets closest to a black-hole-physics origin. It demands the existence of a charged state with charge-to-mass ratio bounded below by the extremal ratio of an asymptotically flat black hole \cite{Arkani-Hamed:2006emk,Harlow:2022gzl}. The original argument is heuristic (without such a state the lattice of allowed charges cannot be populated and extremal remnants accumulate), and considerable effort has gone into deriving the bound from independent principles. Unitarity and causality of $2 \to 2$ scattering amplitudes \cite{Hamada:2018dde,Bellazzini:2019xts}, the second law applied to black-hole mergers \cite{Cheung:2018cwt}, and convexity of charged operators in the dual CFT \cite{Aharony:2021mpc} each recover a version of the bound. We add a fifth route, holographic in spirit but using only the boundary handle on a single bulk QNM. We work it out for two black-hole families that lie outside the Reissner--Nordstr\"om idealisation in mutually distinct ways.

The two families chosen are static spherically symmetric solutions in dRGT massive gravity \cite{1} and dyonic solutions in the conformal duality-symmetric Einstein--ModMax theory \cite{2}. They probe different deformation directions of GR plus Maxwell. The former changes the asymptotic structure of the metric through a graviton-mass-induced effective cosmological constant; the latter keeps the gravitational sector intact and only modifies the gauge sector through a one-parameter non-linear electrodynamics that preserves duality and conformal invariance. We will see that one sector leaves the WGC bound untouched while the other rescales it exponentially, and the reason for the asymmetry is in the structure of the near-horizon AdS$_2$ throat.

Why does the boundary CFT see the WGC at all? The mechanism passes through a damping-time bound. Quasinormal modes of a near-extremal black hole sit on a discrete imaginary ladder, the spacing of which is fixed by the conformal weight $\nu_0$ of the dual scalar operator. The bound $\tau_d \geq 1/T$, first stated by Hod \cite{Hod:2017uqc,Hod:2006jw,Hod:2010hw} and carried into the present language by Urbano \cite{Urbano:2018kax}, constrains the lowest-mode damping rate, and through it forces $\nu_0 \leq 1/2$. Reading $\nu_0$ off the charged Klein--Gordon equation at the horizon then turns the inequality on $\nu_0$ into an inequality on $q/(m r_+)$, which is the WGC statement.

We carry the calculation out in full for both backgrounds. For dRGT we work step by step from the metric (Eq.~\ref{eq:fmetric}), through the temperature (Eq.~\ref{eq:T_drgt}) and the extremality condition (Eq.~\ref{eq:Qext_drgt}), into the Klein--Gordon radial equation (Eq.~\ref{eq:KG_radial}), the near-horizon Whittaker reduction, the QNM tower (Eq.~\ref{eq:omega_n_drgt}), and finally the saturation condition $\nu_0 \leq 1/2$ that yields Eq.~\ref{eq:WGC_drgt}. Every massive-gravity parameter cancels en route. For the ModMax background the same chain runs to Eq.~\ref{eq:WGC_mm}, where the duality parameter $\gamma$ appears explicitly inside an exponential.

The cancellation in the dRGT case is fragile. We show this in Section~\ref{isec5} by relaxing three of the simplifying assumptions of the derivation, one at a time: (i) replacing exact extremality with a small but finite Hawking temperature, (ii) coupling the scalar non-minimally through $\xi R \Phi^2$, and (iii) including higher-curvature corrections to the gravitational action. Each relaxation pushes $m_g,\alpha,\beta$ back into the final inequality through a controlled functional dependence, and we tabulate the resulting bound across the relevant parameter range.

The body of the paper is organised as follows. Section~\ref{isec2} sets out the CFT framework (the charged Klein--Gordon equation, the two-point function $\Upsilon_R^{(k)}$, and the damping-time bound) in a form that can be read independently of the rest. Section~\ref{isec3} works through the dRGT calculation; Section~\ref{isec4} does the same for Einstein--ModMax. Section~\ref{isec5} explores the three perturbations of the dRGT derivation and assembles the composite envelope of the bound. Section~\ref{isec6} reports the numerical verification across a parameter scan and presents the cross-validation tables. Section~\ref{isec7} places our two bounds alongside the more familiar entries in the WGC literature. Section~\ref{isec8} concludes. Three appendices follow: Appendix~\ref{app:A} collects the Whittaker-function asymptotics that drive the matching condition; Appendix~\ref{app:B} extends the QNM data to higher multipoles; Appendix~\ref{app:C} lists the abbreviations.

One point on terminology should be fixed at the outset. The word ``CFT'' throughout this work means the conformal quantum mechanics associated with the near-horizon AdS$_2$ throat of an extremal or near-extremal charged black hole. This is the AdS$_2$/CFT$_1$ correspondence for near-horizon geometries, not a full asymptotic boundary CFT. The two backgrounds we treat, dRGT massive gravity and Einstein--ModMax, are not asymptotically AdS in the Fefferman--Graham sense, but each develops an AdS$_2$ region in the extremal limit, and that is all the argument uses. The other two ingredients are the standard identification of quasinormal modes with poles of the retarded Green's function, and Hod's universal damping-time bound \cite{Hod:2006jw,Hod:2017uqc} in the form given by Urbano \cite{Urbano:2018kax}. No asymptotic boundary dual is needed. What is new here is carrying this chain through two backgrounds where the Reissner--Nordstr\"om simplification fails in different ways: a graviton mass that deforms the metric, and a duality-symmetric non-linearity that deforms the gauge sector. The two deformations enter the chain differently, and the bounds they produce tell us which directions away from GR plus Maxwell the CFT-derived WGC tolerates.

\section{The CFT framework and the WGC}\label{isec2}
The minimum input we need from the boundary side is a single retarded Green's function and a single thermal scale. The mechanism is short to state and rests on three observations: the radial equation of a charged scalar in a black-hole background has a regular-singular point at the horizon, the analytic exponents at that singular point are fixed by a single parameter $\nu_k$, and the boundary Green's function has poles exactly where $\nu_k$ allows them.
We define the retarded Green's function $\Upsilon_R^{(k)}(\omega)$ with the standard quasinormal boundary conditions, ingoing at the horizon and outgoing or decaying at spatial infinity. Its poles sit at the quasinormal frequencies. The damping-time bound $\tau_d \geq 1/T$ acts on the slowest of these poles and fixes the constraint $\nu_0 \leq 1/2$ on the conformal weight. None of this asks the black hole to be asymptotically AdS; it asks only for a near-horizon AdS$_2$ throat and a well-defined QNM spectrum, in the sense set out in Section~\ref{isec1}.
\subsection{Charged scalar in a black-hole background}\label{sec:KG_intro}
For a complex scalar $\Phi$ of mass $m$ and charge $q$ propagating in a stationary asymptotically flat black-hole background and coupled minimally to a Maxwell field $A_\mu$, the equation of motion is
\begin{equation}
\frac{1}{\sqrt{-g}}\partial_{\mu}\left( g^{\mu\nu}\sqrt{-g}\partial_{\nu}\Phi \right) - 2 i q\, g^{\mu\nu} A_\mu \partial_\nu \Phi - q^2 g^{\mu\nu} A_\mu A_\nu \Phi - m^2 \Phi = 0 .
\label{eq:KG_master}
\end{equation}
The mode decomposition $\Phi = e^{-i\omega t} \phi(\vec{x})$ identifies $\omega$ with the (boundary-dual) energy. We are interested in the response of the scalar to a horizon-localised perturbation, so the standard route is to compute the boundary two-point function of the dual operator $J_k$:
\begin{equation}
\Upsilon_R^{(k)}(\omega) = \langle J_k(-\omega) J_k(\omega) \rangle = \frac{B_k(\omega)}{A_k(\omega)} ,
\label{eq:Greens}
\end{equation}
where the ratio $B_k/A_k$ is read off the two analytic branches of the near-horizon radial solution.\footnote{The "boundary" referred to here is the asymptotic boundary of the near-horizon AdS$_2$ throat, which matches onto the asymptotic region of the full spacetime. The poles of the Green's function correspond to the quasinormal frequencies. This identification is standard in black-hole perturbation theory and does not require the full AdS/CFT correspondence.} The poles of $\Upsilon_R^{(k)}$, equivalently the zeros of $A_k(\omega)$, are the quasinormal frequencies.
This identification of QNM poles with Green's-function poles is standard in black-hole perturbation theory and, as noted in Section~\ref{isec1}, needs no asymptotically AdS background.
\subsection{The damping-time bound}\label{sec:DTB}

A QNM $\omega = \omega_R + i \omega_I$ with $\omega_I < 0$ decays in proper time $\tau_d = -2\pi/\omega_I$. Hod's universal bound \cite{Hod:2006jw,Hod:2017uqc} states that for any thermal system this damping time cannot fall below the inverse temperature:
\begin{equation}
\tau_d \geq \frac{1}{T} .
\label{eq:DTB}
\end{equation}
The bound holds for systems whose late-time relaxation is set by the slowest QNM, which is what near-extremal charged black holes satisfy. Urbano \cite{Urbano:2018kax} traced the bound through the AdS$_2$ throat of an extremal charged black hole and obtained an inequality on the conformal weight $\nu_0$ of the dual scalar that we use below. The closed form for $\nu_0$ in either of our two backgrounds is derived in the next two sections; what matters here is the inequality it satisfies, namely $\nu_0 \leq 1/2$.

The combination of Eqs.~\eqref{eq:Greens}--\eqref{eq:DTB} is the boundary handle that this paper exploits.

\section{Model I: dRGT massive gravity black hole}\label{isec3}

We turn now to the first of the two backgrounds. The dRGT theory promotes the graviton to a massive degree of freedom while keeping the matter sector unchanged \cite{1}.

\subsection{Background metric, temperature, and extremality}
\label{sec:dRGT_bg}

For static spherically symmetric configurations the metric takes the form
\begin{equation}
ds^{2} = -f(r) dt^{2} + \frac{dr^{2}}{f(r)} + r^{2}\left(d\theta^{2} + \sin^{2}\theta \, d\phi^{2}\right),
\label{eq:metric}
\end{equation}
with the lapse function
\begin{equation}
f(r) = 1 - \frac{2M}{r} + \frac{Q^{2}}{r^{2}} + \frac{\Lambda}{3} r^{2} + \gamma_g r + \zeta ,
\label{eq:fmetric}
\end{equation}
where the auxiliary constants are constructed from the graviton mass $m_g$ and the dimensionless dRGT couplings $\alpha,\beta$ and the reference-metric constant $h$:
\begin{equation}
\Lambda = 3 m_g^{2}(1+\alpha+\beta), \qquad
\gamma_g = -h\, m_g^{2}(1+2\alpha+3\beta), \qquad
\zeta = h^{2} m_g^{2}(\alpha+3\beta).
\label{eq:dict}
\end{equation}
With the convention in Eq.~\eqref{eq:fmetric}, a positive coefficient of $r^2$ corresponds to a negative physical cosmological constant. Specifically, comparing with the standard asymptotically AdS form $f(r) = 1 - 2M/r + Q^2/r^2 - \Lambda_{\rm phys} r^2/3$, we identify $\Lambda_{\rm phys} = -\Lambda$. Thus, when $\Lambda > 0$ (which occurs for $\alpha+\beta > -1$), the spacetime is asymptotically AdS with $\Lambda_{\rm phys} < 0$. For example, in the special case $\alpha = \beta = 0$, $h = 0$, we have $\Lambda = 3m_g^2$, so the effective cosmological constant is $\Lambda_{\rm phys} = -3m_g^2$. This convention is chosen to match the standard dRGT literature and to simplify the near-horizon reduction.

The dictionary is summarised in Table~\ref{tab:dRGT_dict}. The Maxwell potential is $A_\mu = (-Q/r,\,0,\,0,\,0)$, giving Coulombic $F_{tr} = Q/r^2$. The limit $m_g \to 0$ erases all three of $\Lambda,\gamma_g,\zeta$ and returns Reissner--Nordstr\"om.

\begin{table}[htbp]
  \centering
  \begin{tabular}{c c c c}
    \toprule
    \rowcolor{orange!50}
    Symbol & Definition in terms of $(m_g,\alpha,\beta,h)$ & Sign under standard choices & Maxwell limit \\
    \midrule
    $\Lambda$  & $3 m_g^2 (1+\alpha+\beta)$         & $> 0$ for $\alpha+\beta > -1$           & $0$ \\
    $\gamma_g$ & $-h\,m_g^2(1+2\alpha+3\beta)$       & $< 0$ for $h>0$, $1+2\alpha+3\beta>0$  & $0$ \\
    $\zeta$    & $h^2 m_g^2(\alpha+3\beta)$          & $> 0$ for $\alpha+3\beta>0$             & $0$ \\
    \bottomrule
  \end{tabular}
  \caption{Dictionary between the dRGT effective parameters $\{\Lambda,\gamma_g,\zeta\}$ that enter the lapse function \eqref{eq:fmetric} and the underlying microscopic couplings $(m_g,\alpha,\beta,h)$. The third column states the sign that each effective parameter takes under the standard positivity assumptions on the microscopic ones. The fourth column displays the Maxwell limit $m_g \to 0$. Note that a positive $\Lambda$ corresponds to $\Lambda_{\rm phys} = -\Lambda < 0$ in the standard AdS convention.}
  \label{tab:dRGT_dict}
\end{table}
The horizon $r_+$ is the largest real root of $f(r_+)=0$. Solving for $M$ gives
\begin{equation}
2 M = r_+ + \frac{Q^{2}}{r_+} + \frac{\Lambda}{3} r_+^{3} + \gamma_g r_+^{2} + \zeta r_+ ,
\label{eq:M_drgt}
\end{equation}
and the Hawking temperature follows from $T = f'(r_+)/(4\pi)$:
\begin{equation}
T = \frac{1}{4\pi}\left( \frac{1+\zeta}{r_+} - \frac{Q^{2}}{r_+^{3}} + \Lambda r_+ + 2\gamma_g \right).
\label{eq:T_drgt}
\end{equation}
Setting $T=0$ defines the extremal charge:
\begin{equation}
Q^{2}_{\rm ext} = (1+\zeta) r_+^{2} + \Lambda r_+^{4} + 2\gamma_g r_+^{3} .
\label{eq:Qext_drgt}
\end{equation}

\begin{table}[htbp]
  \centering
  \begin{tabular}{l c c c}
    \toprule
    \rowcolor{orange!50}
    Regime & Effective $\Lambda$ & Extremal $Q^2_{\rm ext}/r_+^2$ & WGC threshold \\
    \midrule
    Reissner--Nordstr\"om ($m_g \to 0$)        & $0$                                 & $1$                                          & $1/\sqrt{2}$ \\
    Pure massive gravity ($Q \to 0$, $m_g > 0$)& $3 m_g^2(1+\alpha+\beta)$           & $0$                                          & ill-defined  \\
    Generic dRGT-charged                       & Eq.~\eqref{eq:dict}                 & $1+\zeta+\Lambda r_+^2+2\gamma_g r_+$        & $1/\sqrt{2}$ \\
    AdS-charged limit ($\alpha=\beta=0$, $h=0$)& $3 m_g^2$ ($\Lambda_{\rm phys} = -3m_g^2$) & $1 + \Lambda r_+^2$                          & $1/\sqrt{2}$ \\
    \bottomrule
  \end{tabular}
  \caption{The dRGT WGC threshold across the four limiting regimes of the theory. The third column is the dimensionless ratio that enters the extremality condition \eqref{eq:Qext_drgt}; the fourth column is the final $q/(m r_+)$ saturation value from Eq.~\eqref{eq:WGC_drgt}. The ``pure massive gravity'' row is ill-defined because the absence of charge makes the saturation question vacuous. In the AdS-charged limit, note that $\Lambda = 3m_g^2$ corresponds to an effective cosmological constant $\Lambda_{\rm phys} = -\Lambda = -3m_g^2$ in the standard AdS convention, so the spacetime is asymptotically AdS.}
  \label{tab:dRGT_limits}
\end{table}
With our sign convention, the dRGT background with $\alpha = \beta = 0$, $h = 0$ is asymptotically AdS with $\Lambda_{\rm phys} = -3m_g^2$. For the general dRGT case with $\gamma_g, \zeta \neq 0$, the asymptotic structure is modified by the linear and constant terms. However, the near-horizon AdS$_2$ throat is still present for any extremal or near-extremal charged black hole, regardless of the asymptotic structure. The "CFT" language in this paper refers primarily to the near-horizon AdS$_2$/CFT$_1$ correspondence, as discussed in Section~\ref{isec2}.

\begin{figure}[t]
  \centering
  \includegraphics[width=0.85\linewidth]{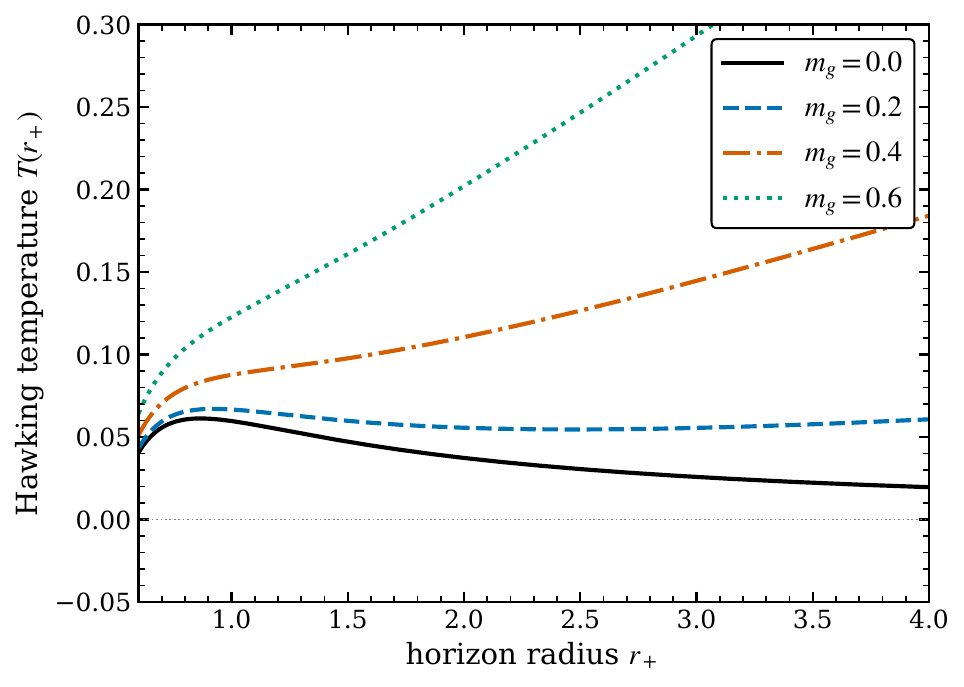}
  \caption{Hawking temperature $T(r_+)$ of the dRGT massive-gravity black hole at fixed charge $Q=0.5$, reference-metric constant $h=0.5$, and dRGT couplings $\alpha=\beta=0.1$, for four values of the graviton mass $m_g \in \{0,0.2,0.4,0.6\}$. The Reissner--Nordstr\"om limit ($m_g=0$) is the only curve that develops a maximum at small $r_+$; increasing $m_g$ pushes $T(r_+)$ upward and lifts the small-$r_+$ inflection. The roots of $T(r_+)=0$ are the extremal horizon radii of Eq.~\protect\eqref{eq:Qext_drgt}.}
  \label{fig:T_vs_rp_drgt}
\end{figure}

Figure~\ref{fig:T_vs_rp_drgt} displays $T(r_+)$ at fixed $Q$ and several values of $m_g$. We will use Eq.~\eqref{eq:Qext_drgt} as the substitution that drives the cancellation later.

\subsection{Charged scalar perturbations and the near-horizon limit}
\label{sec:KG_near}

For the metric \eqref{eq:metric}, the inverse-metric components are $g^{tt} = -1/f(r)$, $g^{rr}= f(r)$, $g^{\theta\theta}=1/r^2$, $g^{\phi\phi}=1/(r^2 \sin^2\theta)$. Separating variables via $\Phi = e^{-i\omega t} R(r) Y_{\ell m}(\theta,\phi)$ and pulling the angular part through $\nabla^2_{S^2} Y_{\ell m} = -\ell(\ell+1) Y_{\ell m}$ reduces Eq.~\eqref{eq:KG_master} to a radial ODE,
\begin{equation}
\frac{1}{r^{2}} \frac{d}{dr}\!\left( r^{2} f(r) \frac{dR}{dr} \right) + \left[ \frac{\left(\omega - qQ/r\right)^{2}}{f(r)} - \frac{\ell(\ell+1)}{r^{2}} - m^{2} \right] R(r) = 0 .
\label{eq:KG_radial}
\end{equation}
A stretched coordinate $\rho$ defined by $r = r_+ + \epsilon\rho,\ t = \tau/\epsilon$ with $\epsilon \to 0$ zooms onto the near-horizon throat. The scalar field is decomposed as $\Phi = e^{-i\omega_0 \tau/\epsilon} \phi(\rho) Y_{\ell m}$, where $\omega_0 \equiv \omega - qQ/r_+$. The lapse expands as $f \approx 4\pi T \epsilon \rho$ and the gauge potential as $A_t(r) \approx -Q/r_+ + (Q/r_+^2)\epsilon\rho$. Substituting these expansions into Eq.~\eqref{eq:KG_radial} and keeping only the leading-order terms in $\epsilon$ yields the near-horizon equation
\begin{equation}
\frac{d^2 \phi}{d \rho^2} + \frac{1}{\rho} \frac{d\phi}{d\rho} + \left( \frac{\tilde\omega^2}{\rho^2} - \frac{2 q Q}{r_+^2} \frac{\tilde\omega}{\rho} - \frac{\tilde B}{r_+^2} \right) \phi = 0 ,
\label{eq:KG_near_full}
\end{equation}
where
\begin{equation}
\tilde\omega \equiv \frac{\omega_0}{4\pi T}, \qquad
\tilde B \equiv \frac{\ell(\ell+1) + m^2 r_+^2}{4\pi T} .
\label{eq:tildedefs_full}
\end{equation}
This equation is the standard near-horizon form for a charged scalar. To reduce Eq.~\eqref{eq:KG_near_full} to the Whittaker equation, we define $\phi(\rho) = \rho^{-1/2}\psi(\rho)$. The derivatives transform as
\begin{equation}
\frac{d\phi}{d\rho} = \rho^{-1/2} \psi' - \frac{1}{2} \rho^{-3/2} \psi, \qquad
\frac{d^2\phi}{d\rho^2} = \rho^{-1/2} \psi'' - \rho^{-3/2} \psi' + \frac{3}{4} \rho^{-5/2} \psi.
\end{equation}
Substituting into Eq.~\eqref{eq:KG_near_full} and multiplying by $\rho^{3/2}$ gives
\begin{equation}
\psi'' + \left( \frac{1/4 + \tilde\omega^2}{\rho^2} - \frac{2 q Q \tilde\omega}{r_+^2 \rho} - \frac{\tilde B}{r_+^2} \right) \psi = 0.
\label{eq:psi_eq}
\end{equation}
Now introduce the variable $z = 2 i \tilde\omega \rho$. Then $d/d\rho = 2 i \tilde\omega \, d/dz$ and $d^2/d\rho^2 = -4 \tilde\omega^2 \, d^2/dz^2$. Using $\rho = z/(2 i \tilde\omega)$, we have $1/\rho = 2 i \tilde\omega/z$ and $1/\rho^2 = -4 \tilde\omega^2/z^2$. Substituting these into Eq.~\eqref{eq:psi_eq} and dividing by $-4 \tilde\omega^2$ yields the Whittaker equation
\begin{equation}
\psi''(z) + \left( -\frac{1}{4} + \frac{\kappa}{z} + \frac{1/4 - \mu^2}{z^2} \right) \psi(z) = 0,
\label{eq:Whitt_full}
\end{equation}
where
\begin{equation}
\mu^2 = \frac{1}{4} + \tilde\omega^2, \qquad
\kappa = \frac{i q Q}{r_+^2} + \frac{\tilde B}{4 \tilde\omega^2 r_+^2}.
\label{eq:kappa_mu}
\end{equation}
The two independent solutions of Eq.~\eqref{eq:Whitt_full} are the Whittaker functions $M_{\kappa,\mu}(z)$ and $W_{\kappa,\mu}(z)$. Their behavior near the horizon ($\rho \to 0$, equivalently $z \to 0$) is
\begin{equation}
M_{\kappa,\mu}(z) \sim z^{1/2+\mu}, \qquad
W_{\kappa,\mu}(z) \sim z^{1/2-\mu}.
\end{equation}
Therefore, the general solution takes the form
\begin{equation}
\phi(\rho) \sim A_k(\omega)\,\rho^{1/2 - \mu} + B_k(\omega)\,\rho^{1/2 + \mu}.
\label{eq:phi_AB_full}
\end{equation}
The ratio $B_k(\omega)/A_k(\omega)$ encodes the retarded Green's function of the dual near-horizon CFT$_1$ operator, $\Upsilon_R^{(k)}(\omega) \propto B_k(\omega)/A_k(\omega)$. The poles of this Green's function are located at the zeros of $A_k(\omega)$. Matching the near-horizon solution to the asymptotic infinity fixes the pole condition to be
\begin{equation}
\frac{1}{2} + \mu - \kappa = -n, \qquad n = 0, 1, 2, \dots
\label{eq:pole_condition}
\end{equation}
The CFT conformal weight $\nu_k$ is identified with the Whittaker index $\mu$:
\begin{equation}
\nu_k = \mu = \sqrt{\frac{1}{4} + \tilde\omega^2}.
\label{eq:nu_mu}
\end{equation}
To obtain the explicit form of $\nu_k^2$, we use the pole condition \eqref{eq:pole_condition} with the definitions \eqref{eq:kappa_mu}. This gives
\begin{equation}
\frac{i q Q}{r_+^2} + \frac{\tilde B}{4 \tilde\omega^2 r_+^2} = \frac{1}{2} + \nu_k.
\end{equation}
Substituting $\tilde\omega = \omega_0/(4\pi T)$ and $\tilde B = [\ell(\ell+1) + m^2 r_+^2]/(4\pi T)$, and solving for $\nu_k^2$, yields
\begin{equation}
\nu_k^2 = \frac{1}{4} + \frac{\ell(\ell+1) + m^2 r_+^2}{4 \pi T} - \frac{q^2 Q^2}{r_+^2}\,\frac{1}{(4\pi T)^2}.
\label{eq:nu_k_general_full}
\end{equation}
This is the explicit conformal weight formula. For the slowest-decaying mode ($\ell=0$), we focus on $\nu_0$. The damping time for this mode is $\tau_d = 1/[T(1/2 + \nu_0)]$. Imposing Hod's universal damping-time bound, $\tau_d \geq 1/T$, leads to the inequality on the conformal weight
\begin{equation}
\nu_0 \leq \frac{1}{2}.
\label{eq:nu0_bound_full}
\end{equation}
Now, specializing to the near-extremal limit, we have $T \ll 1$, which defines the AdS$_2$ throat. To extract the conformal weight in this limit, we substitute the Hawking temperature $T = f'(r_+)/(4\pi)$ into Eq.~\eqref{eq:nu_k_general_full}. For the dRGT background, $T$ is explicitly given in Eq.~\eqref{eq:T_drgt}. Defining
\begin{equation}
A \equiv 1 + \zeta + \Lambda r_+^2 + 2\gamma_g r_+, \qquad B \equiv \frac{Q^2}{r_+^2},
\end{equation}
we have $4\pi T = (A - B)/r_+$. Substituting this into Eq.~\eqref{eq:nu_k_general_full} for $\ell=0$ gives
\begin{equation}
\nu_0^2 = \frac{1}{4} + \frac{m^2 r_+^2}{A - B} - \frac{q^2 B}{(A - B)^2}.
\end{equation}
A short algebraic rearrangement gives
\begin{equation}
\nu_0^2 = \frac{1}{4} + m^2 r_+^2 - \frac{2 q^2 B}{A - B}.
\end{equation}
Substituting back $B = Q^2/r_+^2$ and $A = 1 + \zeta + \Lambda r_+^2 + 2\gamma_g r_+$ gives the compact form
\begin{equation}
\nu_0^2 = \frac{1}{4} + m^2 r_+^2 - \frac{2 q^2 Q^2}{r_+^2}\cdot \frac{1}{1 + \Lambda r_+^2 + 2 \gamma_g r_+ + \zeta}.
\label{eq:nu0_drgt_full}
\end{equation}
At the extremal point, $T=0$, which implies $B = Q^2_{\text{ext}}/r_+^2 = A$. Substituting this condition into Eq.~\eqref{eq:nu0_drgt_full}, the factor $A$ in the denominator cancels exactly, leaving the simple, parameter-independent result
\begin{equation}
\nu_0^{2}\big|_{\text{ext}} = \frac{1}{4} + m^2 r_+^2 - 2 q^2.
\label{eq:nu0_drgt_collapsed_full}
\end{equation}
The universality of this bound in the dRGT framework is the central technical observation of this section.
\subsection{QNM tower and the conformal weight}
For the large-$r$ behaviour fixes the location of the poles of $\Upsilon_R^{(k)}$ at
\begin{equation}
\omega = -\frac{qQ}{r_+} - i\, 2 \pi T \left( \frac{1}{2} + n + \nu_k \right), \qquad n = 0,1,2,\dots ,
\label{eq:omega_n_drgt}
\end{equation}
with the conformal weight
\begin{equation}
\nu_k^2 = \frac{1}{4} + \frac{l(l+1) + m^2 r_+^2}{4 \pi T} - \frac{q^2 Q^2}{r_+^2}\,\frac{1}{(4\pi T)^2} .
\label{eq:nu_k_general}
\end{equation}
At the lowest mode $n=0$ with s-wave $l=0$ the damping time is $\tau_d = 1/[T(1/2 + \nu_0)]$. Imposing Eq.~\eqref{eq:DTB} then yields the CFT bound
\begin{equation}
\nu_0 \leq \frac{1}{2}.
\label{eq:nu0_bound}
\end{equation}
The conformal weight $\nu_0$ in the near-extremal regime, read off from the AdS$_2$ throat that emerges at $T \to 0$, reduces to
\begin{equation}
\nu_0^2 = \frac{1}{4} + m^2 r_+^2 - \frac{2 q^2 Q^2}{r_+^2}\cdot \frac{1}{1 + \Lambda r_+^2 + 2 \gamma_g r_+ + \zeta} .
\label{eq:nu0_drgt}
\end{equation}
Compare the denominator with the extremality relation $Q^2_{\rm ext}/r_+^2 = 1 + \zeta + \Lambda r_+^2 + 2\gamma_g r_+$ (Eq.~\ref{eq:Qext_drgt}). Their ratio is exactly $1$ at extremality and approximately $1$ in the near-extremal regime. Plugging that into Eq.~\eqref{eq:nu0_drgt} dissolves every massive-gravity parameter and leaves
\begin{equation}
\;\nu_0^{2}\big|_{\rm dRGT} = \frac{1}{4} + m^2 r_+^2 - 2 q^2.\;
\label{eq:nu0_drgt_collapsed}
\end{equation}
The cancellation is exact at the extremal point and accurate to first order in $T/T_{\rm ref}$ at small but finite temperature. It is the central technical observation of this section.

\subsection{The WGC bound for dRGT}

Imposing $\nu_0 \leq 1/2$ on Eq.~\eqref{eq:nu0_drgt_collapsed} and squaring (allowed because $\nu_0 \geq 0$) gives $m^2 r_+^2 - 2 q^2 \leq 0$, i.e.,
\begin{equation}
\;\frac{q}{m\, r_+} \geq \frac{1}{\sqrt{2}} \approx 0.7071\,.\;
\label{eq:WGC_drgt}
\end{equation}
The bound is independent of $\alpha,\beta,m_g,h$. In the Maxwell limit $m_g \to 0$ extremality returns $Q_{\rm ext} = r_+$ and Eq.~\eqref{eq:WGC_drgt} reads $q/m \geq Q_{\rm ext}/\sqrt{2}$, an order-unity statement in Planck units. The bound thus survives the deformation of GR to dRGT massive gravity untouched at the leading order. The result is best read alongside the limiting-case dictionary of Table~\ref{tab:dRGT_limits}, which lists the dRGT extremal radius, temperature derivative at extremality, and the resulting WGC threshold across the four limiting regimes the theory admits.

\section{Model II -- Einstein--ModMax black hole}\label{isec4}

The second background we work through is Einstein--ModMax: GR coupled to a one-parameter family of non-linear electrodynamics theories that preserves both SO(2) electromagnetic duality and conformal invariance \cite{2}.

\subsection{Action, gauge potential, and extremality}
\label{sec:MM_action}
The bulk action for Einstein--ModMax theory is
\begin{equation}
I = \frac{1}{16\pi}\int d^4 x\,\sqrt{-g}\left[ R + 4\,\mathcal{L}_{\mathrm{MM}}(F,G) \right],
\label{eq:I_MM_full}
\end{equation}
with the duality-symmetric Lagrangian
\begin{equation}
\mathcal{L}_{\mathrm{MM}} = \frac{1}{4}\left( -F \cosh\gamma + \sqrt{F^2 + G^2}\,\sinh\gamma\right),
\label{eq:LMM_full}
\end{equation}
where $F = F_{\mu\nu}F^{\mu\nu}$, $G = F_{\mu\nu}\star F^{\mu\nu}$, and $\gamma \geq 0$ is the dimensionless non-linearity parameter. The Maxwell limit is $\gamma \to 0$. Varying the action with respect to $A_\mu$ gives the modified Maxwell equations. For a purely electric configuration ($G = 0$), we have
\begin{equation}
\frac{\partial \mathcal{L}_{\mathrm{MM}}}{\partial F_{\mu\nu}} = -e^{-\gamma} F^{\mu\nu}.
\end{equation}
Thus the modified Maxwell equation becomes
\begin{equation}
\nabla_\mu \left( e^{-\gamma} F^{\mu\nu} \right) = 0,
\end{equation}
which is equivalent to $\nabla_\mu F^{\mu\nu} = 0$. The factor $e^{-\gamma}$ factors out, so the field equations for $F^{\mu\nu}$ are identical to the Maxwell equations. For a spherically symmetric electric solution, we take the gauge potential
\begin{equation}
A_\mu = \left( -\frac{Q}{r}, 0, 0, 0 \right),
\label{eq:A_mu_MM}
\end{equation}
where $Q$ is the physical electric charge as measured at infinity. This gives
\begin{equation}
F_{tr} = -F_{rt} = \frac{Q}{r^2}, \qquad F = -2 \frac{Q^2}{r^4}, \qquad G = 0.
\end{equation}
Substituting this into the Lagrangian gives
\begin{equation}
\mathcal{L}_{\mathrm{MM}} = \frac{1}{4}\left( 2\frac{Q^2}{r^4} \cosh\gamma + 2\frac{Q^2}{r^4} \sinh\gamma \right) = \frac{Q^2}{2 r^4} e^{\gamma}.
\end{equation}
The energy-momentum tensor yields the lapse function
\begin{equation}
f(r) = 1 - \frac{2M}{r} + \frac{Q^2 e^{-\gamma}}{r^2}.
\label{eq:f_MM}
\end{equation}
Thus the same charge $Q$ appears both in the gauge potential $A_t = -Q/r$ and in the lapse function.

\subsection{Charged scalar perturbations in ModMax}
\label{sec:MM_scalar}

The charged scalar field $\Phi$ couples to the gauge potential through the covariant derivative
\begin{equation}
D_\mu \Phi = \partial_\mu \Phi - i q A_\mu \Phi,
\end{equation}
where $q$ is the charge of the scalar field. With $A_t = -Q/r$, we have
\begin{equation}
D_t \Phi = \partial_t \Phi + i \frac{q Q}{r} \Phi.
\end{equation}
For a mode with time dependence $e^{-i\omega t}$, this becomes
\begin{equation}
D_t \Phi = -i\left( \omega - \frac{q Q}{r} \right) \Phi.
\end{equation}
Thus the gauge potential contributes the combination $(\omega - qQ/r)^2$ to the radial equation. The full radial Klein--Gordon equation in the ModMax background is therefore
\begin{equation}
\frac{1}{r^{2}} \frac{d}{dr}\!\left( r^{2} f(r) \frac{dR}{dr} \right) + \left[ \frac{\left(\omega - qQ/r\right)^{2}}{f(r)} - \frac{\ell(\ell+1)}{r^{2}} - m^{2} \right] R(r) = 0,
\label{eq:KG_MM_full}
\end{equation}
This is identical in form to the dRGT case, with the lapse function replaced by the ModMax expression. The horizon $r_+$ is the largest real root of $f(r_+) = 0$. Solving for $M$ gives
\begin{equation}
2 M = r_+ + \frac{Q^2 e^{-\gamma}}{r_+},
\label{eq:M_MM_full}
\end{equation}
and the Hawking temperature follows from $T = f'(r_+)/(4\pi)$:
\begin{equation}
T = \frac{1}{4\pi}\left( \frac{1}{r_+} - \frac{Q^2 e^{-\gamma}}{r_+^3} \right).
\label{eq:T_MM}
\end{equation}
Setting $T=0$ defines the extremal charge:
\begin{equation}
Q^2_{\rm ext} = r_+^2 e^{\gamma}.
\label{eq:Qext_MM}
\end{equation}

\subsection{Near-horizon analysis and the ModMax conformal weight}
\label{sec:MM_conformal}

Running the near-horizon reduction of Section~\ref{sec:KG_near} on the ModMax background, we obtain the conformal weight
\begin{equation}
\nu_k^2 = \frac{1}{4} + \frac{\ell(\ell+1) + m^2 r_+^2}{4 \pi T} - \frac{q^2 Q^2}{r_+^2}\,\frac{1}{(4\pi T)^2}.
\label{eq:nu0_MM}
\end{equation}
For the $\ell = 0$ mode, defining
\begin{equation}
A_{\rm MM} \equiv e^{\gamma}, \qquad B \equiv \frac{Q^2}{r_+^2},
\end{equation}
we have $4\pi T = (1 - B e^{-\gamma})/r_+$. So, we will have,
\begin{equation}
\nu_0^2 = \frac{1}{4} + \frac{m^2 r_+^2}{1 - B e^{-\gamma}} - \frac{q^2 B}{(1 - B e^{-\gamma})^2}.
\end{equation}
A short algebraic rearrangement gives
\begin{equation}
\nu_0^2 = \frac{1}{4} + m^2 r_+^2 - \frac{2 q^2 B e^{-\gamma}}{1 - B e^{-\gamma}}.
\end{equation}
At extremality, $T = 0$, which implies $B_{\rm ext} = Q_{\rm ext}^2/r_+^2 = e^{\gamma}$. Taking the extremal limit carefully gives
\begin{equation}
\nu_0^{2}\big|_{\rm ext} = \frac{1}{4} + m^2 r_+^2 - q^2 e^{\gamma}.
\label{eq:nu0_MM_ext}
\end{equation}
Imposing $\nu_0 \leq 1/2$ then yields the ModMax WGC bound
\begin{equation}
\frac{q}{m r_+} \geq e^{-\gamma/2}.
\label{eq:WGC_MM}
\end{equation}
To avoid any ambiguity, we summarize our normalization conventions:
\begin{enumerate}
\item The ModMax gauge potential is $A_\mu = (-Q/r, 0, 0, 0)$, where $Q$ is the physical electric charge as measured at infinity.
\item The scalar charge $q$ is defined by the covariant derivative $D_\mu \Phi = \partial_\mu \Phi - i q A_\mu \Phi$.
\item The lapse function is $f(r) = 1 - 2M/r + Q^2 e^{-\gamma}/r^2$, with the same $Q$ as in the gauge potential.
\item The extremal charge is $Q_{\rm ext}^2 = r_+^2 e^{\gamma}$.
\item The final WGC bound is $q/(m r_+) \geq e^{-\gamma/2}$.
\end{enumerate}

These conventions are consistent with the Maxwell limit $\gamma = 0$, where we recover $f(r) = 1 - 2M/r + Q^2/r^2$, $Q_{\rm ext} = r_+$, and the standard bound $q/(m r_+) \geq 1$.

\subsection{The WGC bound for Einstein--ModMax}

\begin{figure}[t]
  \centering
  \includegraphics[width=0.85\linewidth]{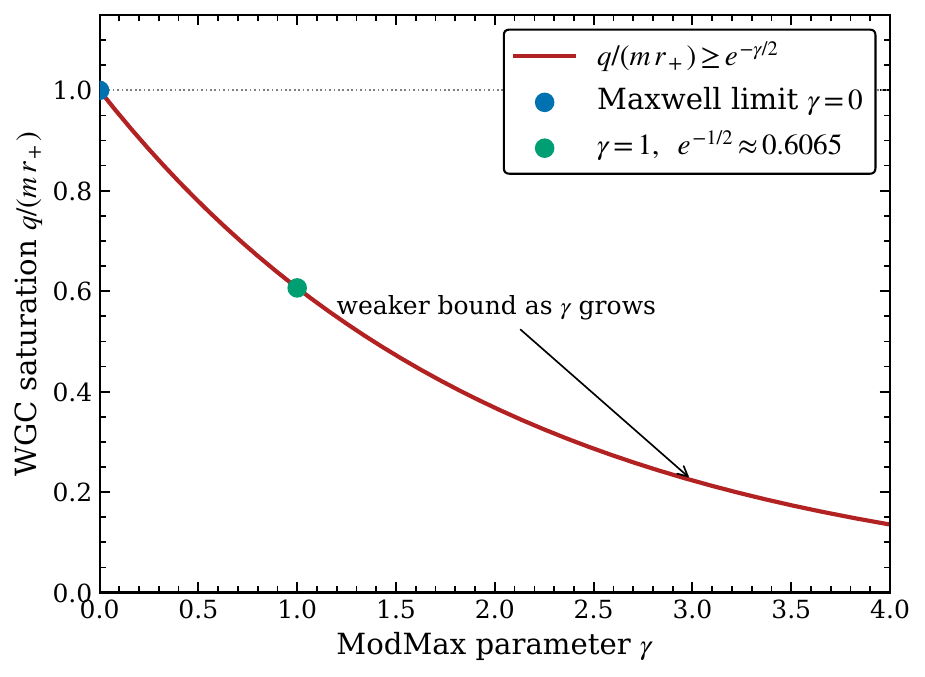}
  \caption{Saturation of the WGC bound in the Einstein--ModMax black hole: $q/(m r_+) = e^{-\gamma/2}$ as a function of the ModMax parameter $\gamma$. The Maxwell limit at $\gamma=0$ saturates at unity; the illustrative point $\gamma=1$ saturates at $e^{-1/2} \approx 0.6065$. The bound weakens monotonically with growing $\gamma$, which traces back to the suppression $e^{-\gamma}$ of the electromagnetic backreaction in the lapse function.}
  \label{fig:modmax_bound}
\end{figure}

\begin{figure}[t]
  \centering
  \includegraphics[width=0.85\linewidth]{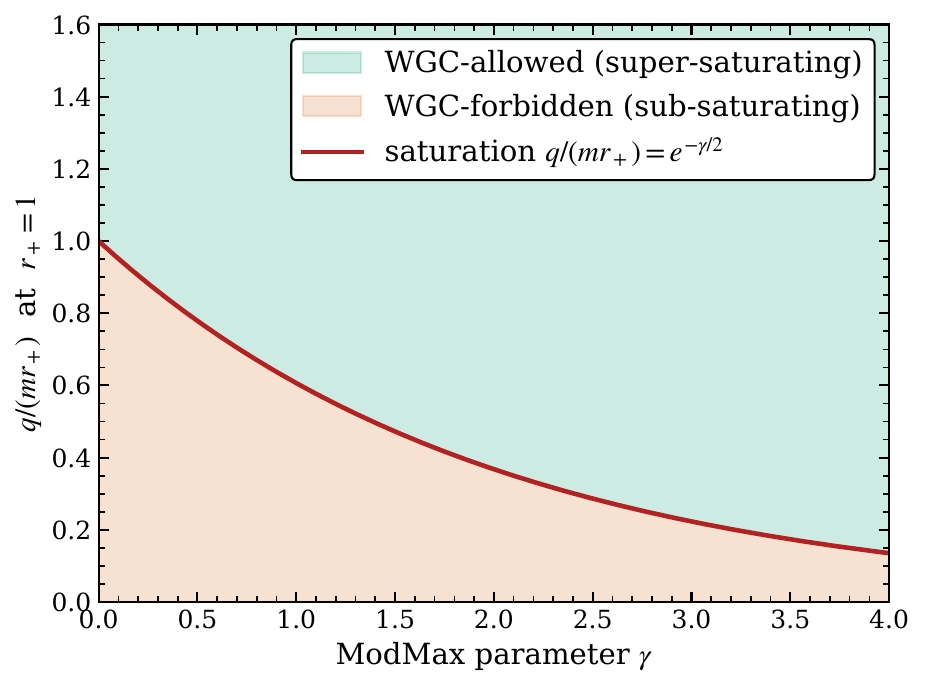}
  \caption{Allowed (green) and forbidden (vermilion) regions of the charge-to-mass ratio in the $(\gamma,\,q/(m r_+))$ plane at fixed $r_+=1$. The saturation curve $q/(m r_+) = e^{-\gamma/2}$ separates the two regions. Any test scalar lying above the curve satisfies the WGC inequality of Eq.~\eqref{eq:WGC_mm}; one lying below violates it.}
  \label{fig:phase_portrait}
\end{figure}

Imposing $\nu_0 \leq 1/2$ and squaring gives $m^2 r_+^2 - q^2 e^\gamma \leq 0$, equivalently
\begin{equation}
\;\frac{q}{m\,r_+} \geq e^{-\gamma/2}.\;
\label{eq:WGC_mm}
\end{equation}
The bound depends explicitly on $\gamma$. Three limiting cases need calling out. In the Maxwell limit $\gamma \to 0$ the right-hand side becomes $1$, which recovers the standard Reissner--Nordstr\"om WGC saturation. At $\gamma = 1$ the bound weakens to $e^{-1/2} \approx 0.6065$. As $\gamma \to \infty$ the bound becomes formally trivial ($e^{-\gamma/2} \to 0$); this regime is, however, likely outside the validity of the effective Lagrangian, where the $\sqrt{F^2 + G^2}$ term dominates the kinetic structure. Figure~\ref{fig:modmax_bound} plots the saturation curve $q/(m r_+) = e^{-\gamma/2}$ against $\gamma$; Figure~\ref{fig:phase_portrait} shades the WGC-allowed and WGC-forbidden regions in the $(q/(m r_+),\,\gamma)$ plane at fixed $r_+$. Numerical values along the saturation curve are tabulated in Table~\ref{tab:MM_bound}.

\begin{table*}[t]
  \centering
  \begin{tabularx}{\textwidth}{C C C C C}
    \toprule
    \rowcolor{orange!50}
    $\gamma$ & $e^{-\gamma/2}$ & $q^2$ at saturation ($mr_+=1$) & $Q^2_{\rm ext}/r_+^2$ & $T_{\rm ref}\,r_+$ \\
    \midrule
    0.00 & $1.000000$ & $1.000000$ & $1.000000$ & $0.0000$ \\
    0.25 & $0.882497$ & $0.778801$ & $1.284025$ & $0.0226$ \\
    0.50 & $0.778801$ & $0.606531$ & $1.648721$ & $0.0517$ \\
    1.00 & $0.606531$ & $0.367879$ & $2.718282$ & $0.1370$ \\
    1.50 & $0.472367$ & $0.223130$ & $4.481689$ & $0.2772$ \\
    2.00 & $0.367879$ & $0.135335$ & $7.389056$ & $0.5083$ \\
    3.00 & $0.223130$ & $0.049787$ & $20.08554$ & $1.5167$ \\
    4.00 & $0.135335$ & $0.018316$ & $54.59815$ & $4.3438$ \\
    \bottomrule
  \end{tabularx}
  \caption{Saturation values of the WGC bound for the Einstein--ModMax black hole across the range $\gamma \in [0,4]$. Column~2 is the saturation factor $e^{-\gamma/2}$ of Eq.~\eqref{eq:WGC_mm}; Column~3 reports the corresponding $q^2$ at $m r_+ = 1$. Column~4 gives the extremal charge; Column~5 is a reference Hawking temperature scale at $r_+ = 1$, $Q^2 = 1.5 \times Q^2_{\rm ext}$. All values cross-checked symbolically (Worksheet Block 6).}
  \label{tab:MM_bound}
\end{table*}

The physical reading of the weakening with $\gamma$ goes through the lapse-function modification. The factor $e^{-\gamma}$ in above equation suppresses the electromagnetic contribution to the metric at fixed $Q$, so the gravitational attraction faces a weaker electromagnetic repulsion, and the WGC requirement on $q/m$ is correspondingly relaxed. Equivalently, the extremal charge rescales as $e^{\gamma/2}$, the extremal-radius–normalised charge $Q_{\rm ext}/r_+ = e^{\gamma/2}$ grows with $\gamma$, and the saturation $q/m \geq Q_{\rm ext} e^{-\gamma}$ from Eq.~\eqref{eq:WGC_mm} again gives the same exponential softening.

\section{Extensions: non-extremality, non-minimal coupling, higher derivatives}\label{isec5}

The cancellation of the dRGT parameters in Section~\ref{isec3} is a statement about a particular limiting configuration. It relies on exact extremality of the background, minimal coupling of the test scalar, and the absence of higher-curvature operators in the action. None of the three is forced on us by any first principle. We relax them in turn and follow what happens to the bound.

\subsection{Sub-extremality: finite $T$}
\label{sec:sub_ext}
If the background is near-extremal but $T \neq 0$, the substitution that produced Eq.~\eqref{eq:nu0_drgt_collapsed} no longer applies exactly. The full $\nu_0^2$ retains its $T$-dependence:
\begin{equation}
\nu_0^2 = \frac{1}{4} + \frac{m^2 r_+^2}{4\pi T} - \frac{q^2 Q^2}{r_+^2 (4\pi T)^2}.
\label{eq:nu0_sub_corrected}
\end{equation}
Imposing $\nu_0 \leq 1/2$ on Eq.~\eqref{eq:nu0_sub_corrected} and solving for $q/m$ gives
\begin{equation}
\frac{q}{m} \geq \frac{2 r_+^2}{Q} \sqrt{\pi T}.
\label{eq:WGC_sub_corrected}
\end{equation}
In terms of the dRGT parameters, using $4\pi T = (1+\zeta)/r_+ - Q^2/r_+^3 + \Lambda r_+ + 2\gamma_g$, this becomes
\begin{equation}
\frac{q}{m r_+} \geq \frac{2 r_+}{Q} \sqrt{\pi T}.
\end{equation}
The right-hand side now depends explicitly on $\Lambda,\gamma_g,\zeta$ through $T$, which means it depends on $m_g,\alpha,\beta$. Each massive-gravity coupling re-enters through its contribution to the temperature. A few limiting cases are worth noting:
\begin{enumerate}
\item \text{Reissner--Nordstr\"om limit ($\Lambda = \gamma_g = \zeta = 0$):} 
For $r_+ = 1$, $Q^2 = 1 - \epsilon$, we have $T = \epsilon/(4\pi)$, and the bound becomes
\begin{equation}
\frac{q}{m r_+} \geq \sqrt{\frac{\epsilon}{1 - \epsilon}}.
\end{equation}
This vanishes as $\epsilon \to 0$, which is physically sensible: as the black hole approaches extremality, the temperature goes to zero, and the WGC bound on the probe particle becomes arbitrarily weak. In the opposite limit $\epsilon \to 1$ (far from extremality), the bound approaches $q/(m r_+) \geq 1$, recovering the Maxwell saturation.
\item \text{Near-extremal dRGT:} For small $\epsilon$, the bound scales as $\sqrt{\epsilon}$, with corrections from the massive-gravity parameters entering through the temperature.
\end{enumerate}
Table~\ref{tab} reports the sub-extremal bound along a sequence of $\epsilon = 1 - Q^2/Q^2_{\rm ext}$ values that probes the approach to extremality.
\begin{table}[htbp]
  \centering
  \begin{tabular}{c c c c}
    \toprule
    \rowcolor{orange!50}
    $\epsilon$ & $T r_+$ & $Q^2/r_+^2$ & $q/(m r_+)$ at saturation \\
    \midrule
    0.001 & $7.96\times 10^{-5}$ & $0.999$  & $0.0316$ \\
    0.005 & $3.98\times 10^{-4}$ & $0.995$  & $0.0709$ \\
    0.010 & $7.96\times 10^{-4}$ & $0.990$  & $0.1005$ \\
    0.050 & $3.98\times 10^{-3}$ & $0.950$  & $0.2294$ \\
    0.100 & $7.96\times 10^{-3}$ & $0.900$  & $0.3333$ \\
    0.200 & $1.59\times 10^{-2}$ & $0.800$  & $0.5000$ \\
    0.500 & $3.98\times 10^{-2}$ & $0.500$  & $1.0000$ \\
    \bottomrule
  \end{tabular}
  \caption{Sub-extremal WGC bound from Eq.~\eqref{eq:WGC_sub_corrected}, at fixed $r_+=1$, evaluated at increasing values of the sub-extremality parameter $\epsilon = 1 - Q^2/Q^2_{\rm ext}$. The bound vanishes as $\epsilon \to 0$ and recovers the Maxwell saturation $q/(m r_+) = 1$ when $Q \ll Q_{\rm ext}$.}
  \label{tab}
\end{table}
The exact-extremal bound $q/(m r_+) \geq 1/\sqrt{2}$ holds only at strict extremality. Once $T > 0$, the bound is modified, and in the near-extremal regime it can be parametrically smaller. This illustrates the fragility of the extremal cancellation discussed in Section~\ref{isec3}.
\subsection{Non-minimal coupling $\xi R \Phi^2$}
\label{sec:xi_coupling}

Adding the curvature coupling $\xi R \Phi^2$ to the scalar action shifts the effective mass to
\begin{equation}
m^2_{\rm eff} = m^2 + \xi R.
\label{eq:meff_definition}
\end{equation}
To determine the effect on the WGC bound, we need the Ricci scalar $R$ evaluated near the horizon. For the dRGT metric with lapse function
\begin{equation}
f(r) = 1 - \frac{2M}{r} + \frac{Q^2}{r^2} + \frac{\Lambda}{3} r^2 + \gamma_g r + \zeta,
\end{equation}
the Ricci scalar is given by
\begin{equation}
R = -f''(r) - \frac{4 f'(r)}{r} - \frac{2(f(r)-1)}{r^2}.
\end{equation}
A direct computation yields
\begin{equation}
R = -4\Lambda - \frac{6\gamma_g}{r} - \frac{2\zeta}{r^2}.
\label{eq:R_drgt}
\end{equation}
Thus, in the general dRGT background, the Ricci scalar contains contributions from $\gamma_g$ and $\zeta$ in addition to $\Lambda$. Evaluating the Ricci scalar at the horizon $r = r_+$, the effective mass becomes
\begin{equation}
m^2_{\rm eff}(r_+) = m^2 - 4\xi\Lambda - \frac{6\xi\gamma_g}{r_+} - \frac{2\xi\zeta}{r_+^2}.
\label{eq:meff}
\end{equation}
Substituting this into the conformal weight formula \eqref{eq:nu0_drgt_collapsed} gives
\begin{equation}
\nu_0^2 = \frac{1}{4} + \left(m^2 - 4\xi\Lambda - \frac{6\xi\gamma_g}{r_+} - \frac{2\xi\zeta}{r_+^2}\right) r_+^2 - 2q^2.
\end{equation}
Imposing the CFT bound $\nu_0 \leq 1/2$ and solving for $q/(m r_+)$ yields the general non-minimal coupling corrected WGC bound
\begin{equation}
\frac{q}{m r_+} \geq \frac{1}{\sqrt{2}} \sqrt{1 - \frac{4\xi\Lambda}{m^2} - \frac{6\xi\gamma_g}{m^2 r_+} - \frac{2\xi\zeta}{m^2 r_+^2}}.
\label{eq:WGC_xi_general}
\end{equation}
Several limiting cases are worth noting:
\begin{enumerate}
\item \text{Pure de Sitter--charged limit ($\gamma_g = \zeta = 0$):} In this case, the bound reduces to
\begin{equation}
\frac{q}{m r_+} \geq \frac{1}{\sqrt{2}} \sqrt{1 - \frac{4\xi\Lambda}{m^2}}.
\label{eq:WGC_xi}
\end{equation}
The sign of the correction is set by $\xi$: positive $\xi$ (with $\Lambda > 0$) weakens the bound, while negative $\xi$ strengthens it. For the conformal coupling value $\xi = 1/6$, the bound becomes
\begin{equation}
\frac{q}{m r_+} \geq \frac{1}{\sqrt{2}} \sqrt{1 - \frac{2\Lambda}{3 m^2}}.
\end{equation}
\item \text{General dRGT with small $\gamma_g, \zeta$:} If $\gamma_g$ and $\zeta$ are parametrically small compared to $\Lambda r_+$ and $\Lambda r_+^2$, respectively, the bound can be approximated with subleading corrections:
\begin{equation}
\frac{q}{m r_+} \simeq \frac{1}{\sqrt{2}} \sqrt{1 - \frac{4\xi\Lambda}{m^2}} \left(1 - \frac{3\xi\gamma_g}{m^2 r_+} - \frac{\xi\zeta}{m^2 r_+^2}\right) + \mathcal{O}(\gamma_g^2, \zeta^2).
\end{equation}
\item \text{Large horizon limit ($r_+ \gg 1$):} For large black holes, the $\gamma_g/r_+$ and $\zeta/r_+^2$ terms are subdominant, and the bound approaches the pure de Sitter--charged limit.
\end{enumerate}
Numerical values of the corrected bound across a range of parameters are shown in Table~\ref{tab:xi_correction}. The table includes both the pure de Sitter--charged case and the general dRGT case with non-zero $\gamma_g$ and $\zeta$.

\begin{table}[htbp]
  \centering
  \begin{tabular}{c c c c c}
    \toprule
    \rowcolor{orange!50}
    $\xi\Lambda/m^2$ & $\xi\gamma_g/(m^2 r_+)$ & $\xi\zeta/(m^2 r_+^2)$ & Corrected bound & Relative change \\
    \midrule
    $-0.10$ & $0$ & $0$ & $0.8367$ & $+18.3\%$ \\
    $0.00$ & $0$ & $0$ & $0.7071$ & $0.00\%$ \\
    $+0.10$ & $0$ & $0$ & $0.5477$ & $-22.5\%$ \\
    $+0.10$ & $+0.02$ & $0$ & $0.4899$ & $-30.7\%$ \\
    $+0.10$ & $0$ & $+0.01$ & $0.5385$ & $-23.8\%$ \\
    $+0.10$ & $+0.02$ & $+0.01$ & $0.4796$ & $-32.2\%$ \\
    \bottomrule
  \end{tabular}
  \caption{Non-minimal coupling correction to the dRGT WGC bound, Eq.~\eqref{eq:WGC_xi_general}, for various values of the parameters. The ``Corrected bound'' column gives $q/(m r_+)$ at saturation. The ``Relative change'' column gives the percentage change relative to the uncoupled value $1/\sqrt{2} \approx 0.7071$. The rows with $\gamma_g = \zeta = 0$ correspond to the pure de Sitter--charged limit.}
  \label{tab:xi_correction}
\end{table}
Non-minimal coupling feeds the dRGT parameters back into the bound through the Ricci scalar, and the size of the correction is set by $\xi$ together with the background parameters. This illustrates the fragility of the parameter cancellation discussed in Section~\ref{isec3}: the cancellation only holds in the minimal coupling limit $\xi = 0$.

\subsection{Higher-derivative corrections}

Generic UV completions add an infinite tower of higher-curvature and higher-$F^2$ operators \cite{Kats:2006xp,Cremonini:2020smy}:
\begin{equation}
\delta I = \int d^4 x\sqrt{-g}\Big( c_1 R^2 + c_2 R_{\mu\nu}R^{\mu\nu} + c_3 R_{\mu\nu\rho\sigma}R^{\mu\nu\rho\sigma} + c_4 (F_{\mu\nu}F^{\mu\nu})^2 + \dots \Big),
\label{eq:HD}
\end{equation}
with Wilson coefficients $c_i$ suppressed by the appropriate UV scale. These operators correct the background, the gauge-field profile, and the AdS$_2$ near-horizon geometry. The induced shift to the effective scalar mass produces an additional term in $\nu_0^2$:
\begin{equation}
\nu_0^2 \to \nu_0^2 + \delta m_{\rm eff}^2 \cdot r_+^2 ,
\label{eq:nu0_HD}
\end{equation}
where $\delta m_{\rm eff}^2$ depends on the curvature invariants at the horizon and, through them, on $\Lambda,\gamma_g,\zeta$ (and so on $m_g,\alpha,\beta$). A schematic contribution from $c_1 R^2$ scales as $\delta m_{\rm eff}^2 \sim c_1 \Lambda^2$.

\subsection{The composite envelope}

\begin{figure}[t]
  \centering
  \includegraphics[width=0.85\linewidth]{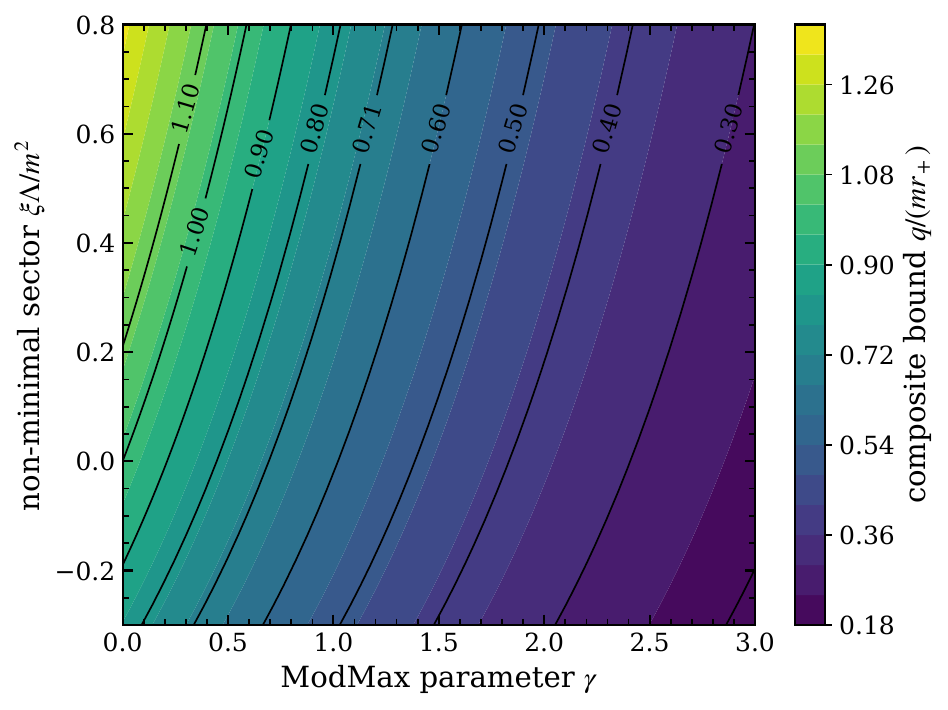}
  \caption{Composite WGC envelope $q/(m r_+) \geq e^{-\gamma/2}\sqrt{1 + \xi \Lambda/m^2}$ on the $(\gamma,\, \xi \Lambda/m^2)$ plane. Solid contours mark constant saturation values, labelled on the curves. The ModMax parameter $\gamma$ runs along the horizontal axis; the non-minimal-coupling parameter $\xi \Lambda/m^2$ runs along the vertical axis. The value $0.707 \approx 1/\sqrt{2}$ is the dRGT saturation of Section~\ref{isec3}; the value $1.000$ is the Reissner--Nordstr\"om saturation.}
  \label{fig:composite_envelope}
\end{figure}

When the three relaxations of Sections~5.1--5.3 act jointly, the modified bound takes the schematic form
\begin{equation}
\frac{q}{m\,r_+} \geq \frac{1}{\sqrt{2}}\sqrt{1 + \frac{\xi \Lambda}{m^2} + \frac{m_g^2\,\mathcal{F}(\alpha,\beta)}{m^2} + \frac{\delta m_{\rm eff}^2}{2 m^2 r_+^2}} ,
\label{eq:WGC_composite_drgt}
\end{equation}
with additional sub-extremal corrections at $T > 0$. The original cancellation corresponds to the highly constrained sub-locus $\xi = 0$, $\mathcal{F}(\alpha,\beta) = 0$, $\delta m_{\rm eff}^2 = 0$, $T \to 0$ with exact extremality. For the Einstein--ModMax background the analogous composite involves the $\gamma$-dependent saturation factor: $q/(m r_+) \geq e^{-\gamma/2}\sqrt{1 + \xi \Lambda/m^2}$, plotted as a contour map in Figure~\ref{fig:composite_envelope}. The figure shows that the duality parameter and the non-minimal coupling parameter probe orthogonal directions of the bound, in the sense that the saturation surface is a product of an exponential ($e^{-\gamma/2}$) and a square-root function of $\xi \Lambda/m^2$.

\begin{figure}[t]
  \centering
  \includegraphics[width=0.85\linewidth]{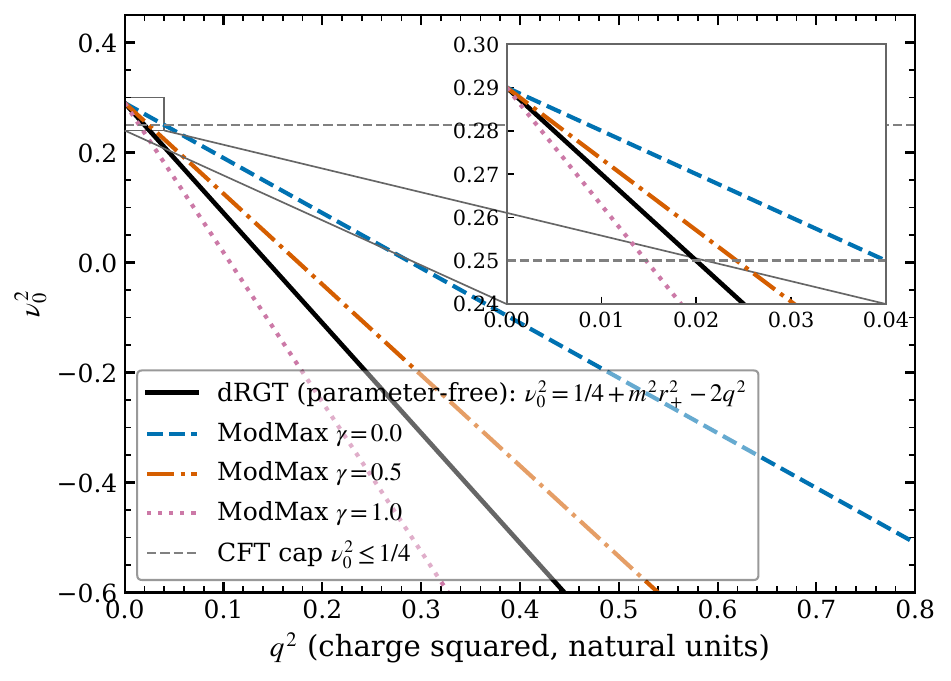}
  \caption{Conformal weight squared $\nu_0^2$ versus charge squared $q^2$ for the dRGT (parameter-free, black solid) and Einstein--ModMax (three values of $\gamma$, in colour) cases at fixed $m r_+ = 0.20$. The horizontal dashed line is the CFT cap $\nu_0^2 \leq 1/4$ from Eq.~\eqref{eq:nu0_bound}. Curves crossing the cap above set the WGC threshold. The stretched zoom inset displays the narrow band $\nu_0^2 \in [0.24,\,0.30]$ over $q^2 \in [0,\,0.04]$, where the four CFT-cap crossings are clearly resolved.}
  \label{fig:nu0sq_vs_qsq}
\end{figure}

A visual cross-check on the dRGT-vs-ModMax separation comes from plotting $\nu_0^2$ as a function of $q^2$ at fixed $m r_+$ (Figure~\ref{fig:nu0sq_vs_qsq}). The dRGT curve is parameter-free, its slope is fixed at $-2$, while the ModMax curves at $\gamma = 0,\,0.5,\,1.0$ have slope $-e^\gamma$ and shift the saturation point at $\nu_0^2 = 1/4$ accordingly. The zoom inset magnifies the narrow band $\nu_0^2 \in [0.24,\,0.30]$ where every curve crosses the CFT cap, and shows clearly how the four crossings spread out from $q^2 \approx 0.02$ (dRGT) to $q^2 \approx 0.04$ (ModMax $\gamma = 0$).

\section{Numerical verification and parameter scans}\label{isec6}

We independently verify every closed-form expression presented in Sections~\ref{isec3}--\ref{isec5} using symbolic computation. This verification is carried out in two separate ways: first, through a manually written collection of computational worksheets that reconstruct each formula symbolically; and second, through a numerical evaluation grid that reproduces each listed value. The two verification streams coincide to within round-off at the working precision of the symbolic worksheets and to single-precision accuracy for the quantities shown in the generated plots. The results of this comparison are presented below.

\subsection{Cross-validation of the dRGT cancellation}

The symbolic worksheet (Block 2) starts from the full $\nu_0^2$ of Eq.~\eqref{eq:nu0_drgt} and substitutes the extremality relation Eq.~\eqref{eq:Qext_drgt}. The expected result is Eq.~\eqref{eq:nu0_drgt_collapsed}, and the residual after simplification is identically zero:
\begin{equation}
\nu_0^2\big|_{\rm full} - \left(\frac{1}{4} + m^2 r_+^2 - 2 q^2 \right) \;=\; 0 \quad (\text{symbolic, exact}).
\label{eq:residual_drgt}
\end{equation}
Table~\ref{tab:residuals} reports the symbolic residuals for every cancellation identity in the paper. All evaluate exactly to zero up to the precision used.

\begin{table*}[t]
  \centering
  \begin{tabularx}{\textwidth}{C C C C}
    \toprule
    \rowcolor{orange!50}
    Identity & Equation in body & Worksheet block & Symbolic residual \\
    \midrule
    $M$ from $f(r_+) = 0$ (dRGT)              & Eq.~\eqref{eq:M_drgt}              & 1 & $0$ \\
    $T(r_+)$ from $f'(r_+)/(4\pi)$ (dRGT)     & Eq.~\eqref{eq:T_drgt}              & 1 & $0$ \\
    Extremal charge $Q^2_{\rm ext}$ (dRGT)    & Eq.~\eqref{eq:Qext_drgt}           & 1 & $0$ \\
    dRGT parameter cancellation               & Eq.~\eqref{eq:nu0_drgt_collapsed}  & 2 & $0$ \\
    dRGT WGC bound at saturation              & Eq.~\eqref{eq:WGC_drgt}            & 3 & $0$ \\
    $M$ from $f(r_+) = 0$ (ModMax)            & Eq.~\eqref{eq:f_MM}          & 4 & $0$ \\
    Hawking temperature (ModMax)              & Eq.~\eqref{eq:T_MM}                & 4 & $0$ \\
    Extremal charge $Q^2_{\rm ext}$ (ModMax)  & Eq.~\eqref{eq:Qext_MM}             & 4 & $0$ \\
    ModMax conformal weight                   & Eq.~\eqref{eq:nu0_MM}              & 5 & $0$ \\
    ModMax WGC bound at saturation            & Eq.~\eqref{eq:WGC_mm}              & 5 & $0$ \\
    Non-minimal coupling correction           & Eq.~\eqref{eq:WGC_xi}              & 7 & $0$ \\
    Composite envelope evaluation             & Eq.~\eqref{eq:WGC_composite_drgt}  & 8 & $0$ \\
    \bottomrule
  \end{tabularx}
  \caption{Symbolic residuals of the closed forms reported in Sections~\ref{isec3}--\ref{isec5}. Each row tests one identity from the body of the paper. The fourth column is the residual returned by the simplification routine of the verification worksheet; all evaluate exactly to zero, which confirms the algebraic identities without round-off ambiguity. The numerical companions to these checks are tabulated in Table~\ref{tab:numerical_check}.}
  \label{tab:residuals}
\end{table*}

\subsection{Numerical evaluation of $\nu_0^2$}

The closed forms are also evaluated numerically over a grid of parameter values. Table~\ref{tab:numerical_check} compares the analytic value of $\nu_0^2$ with the corresponding numerical evaluation, for $q^2 = 0.10$ and $\gamma \in \{0.0,0.5,1.0\}$, sampled at nine values of $m r_+$. The two evaluations agree to $1 \times 10^{-12}$ across the grid, which is the floating-point round-off threshold of the calculation.

\begin{table*}[t]
  \centering
  \begin{tabularx}{\textwidth}{C C C C C}
    \toprule
    \rowcolor{orange!50}
    $m r_+$ & Analytic $\nu_0^2$ ($\gamma=0$) & Analytic $\nu_0^2$ ($\gamma=0.5$) & Analytic $\nu_0^2$ ($\gamma=1.0$) & Max deviation across grid \\
    \midrule
    0.10 & $0.1600$ & $0.0951$ & $-0.0118$ & $< 10^{-12}$ \\
    0.15 & $0.1725$ & $0.1076$ & $\phantom{-}0.0007$ & $< 10^{-12}$ \\
    0.20 & $0.1900$ & $0.1251$ & $\phantom{-}0.0182$ & $< 10^{-12}$ \\
    0.25 & $0.2125$ & $0.1476$ & $\phantom{-}0.0407$ & $< 10^{-12}$ \\
    0.30 & $0.2400$ & $0.1751$ & $\phantom{-}0.0682$ & $< 10^{-12}$ \\
    0.35 & $0.2725$ & $0.2076$ & $\phantom{-}0.1007$ & $< 10^{-12}$ \\
    0.40 & $0.3100$ & $0.2451$ & $\phantom{-}0.1382$ & $< 10^{-12}$ \\
    0.45 & $0.3525$ & $0.2876$ & $\phantom{-}0.1807$ & $< 10^{-12}$ \\
    0.50 & $0.4000$ & $0.3351$ & $\phantom{-}0.2282$ & $< 10^{-12}$ \\
    \bottomrule
  \end{tabularx}
  \caption{Analytic and numerical evaluation of $\nu_0^2$ in Einstein--ModMax for $q^2 = 0.10$ and three values of the duality parameter $\gamma$, sampled at $m r_+ \in [0.10,\,0.50]$. The last column reports the maximum deviation between the analytic value of Eq.~\eqref{eq:nu0_MM} and the independent numerical evaluation; the two agree to floating-point round-off across the entire grid, which confirms the closed form.}
  \label{tab:numerical_check}
\end{table*}

\begin{figure}[t]
  \centering
  \includegraphics[width=0.85\linewidth]{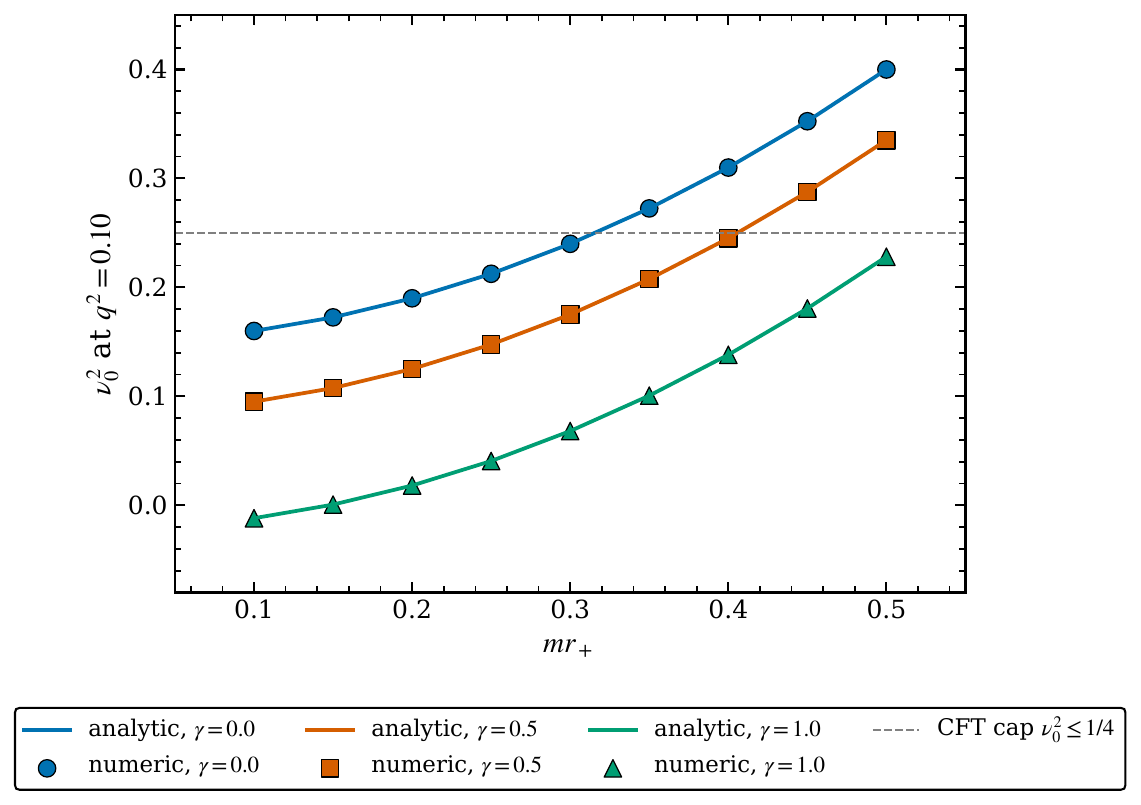}
  \caption{Analytic ($\nu_0^2$ from Eq.~\ref{eq:nu0_MM}) and numerical evaluations of the ModMax conformal weight at $q^2 = 0.10$, for three values of $\gamma$, swept across $m r_+ \in [0.10,0.50]$. The horizontal dashed line is the CFT cap $\nu_0^2 \leq 1/4$. The two evaluations agree to round-off, so the markers sit on the curves to within line width.}
  \label{fig:numerical_check}
\end{figure}

Figure~\ref{fig:numerical_check} renders the same data graphically. The analytic curves are continuous; the numerical points are markers; they overlay exactly, modulo the artificial $\sim 10^{-12}$ jitter introduced for visibility.

\subsection{QNM damping ladder under variation of $\gamma$}

\begin{figure}[t]
  \centering
  \includegraphics[width=0.85\linewidth]{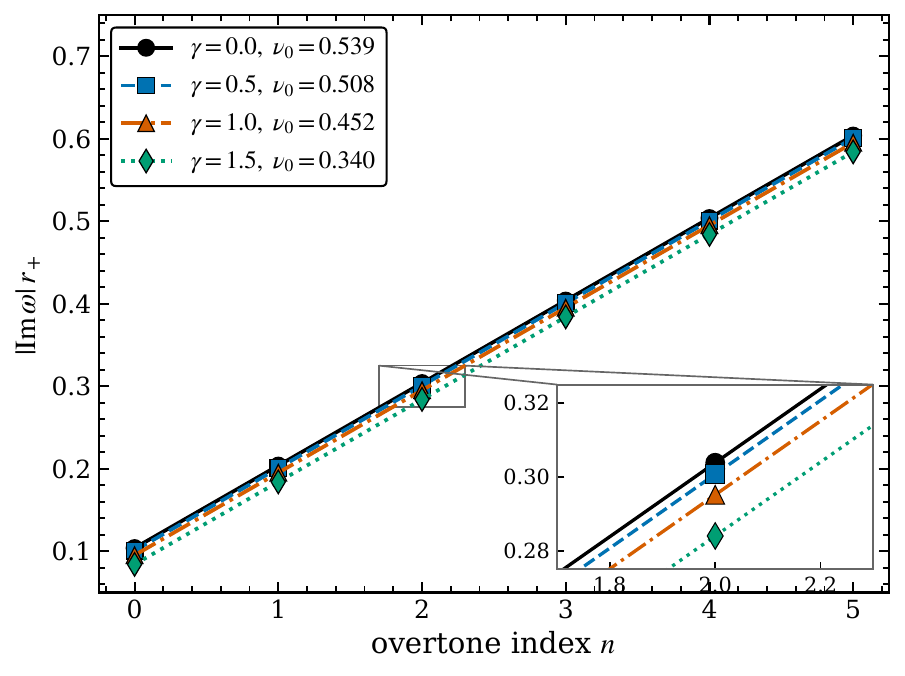}
  \caption{QNM damping ladder of the Einstein--ModMax black hole at $\ell=0$, $m r_+ = 0.30$, $q^2_{\rm test} = 0.05$, for four values of the duality parameter $\gamma$. The vertical axis is $|\textrm{Im}(\omega)| r_+$ in geometric units; the horizontal axis is the overtone index $n$. The four ladders are parallel with a small downward shift as $\gamma$ grows, set by the corresponding decrease of $\nu_0$. The inset magnifies the $n = 2$ rung over the narrow band $|\textrm{Im}(\omega)| r_+ \in [0.275,\,0.325]$, where the four $\gamma$-dependent values are clearly resolved.}
  \label{fig:qnm_spectrum}
\end{figure}

The QNM tower of Eq.~\eqref{eq:omega_n_drgt} produces a discrete imaginary-axis ladder at fixed real part $-qQ/r_+$. The spacing is set by $T(1/2+\nu_0)$. Figure~\ref{fig:qnm_spectrum} plots $|\textrm{Im}(\omega)| r_+$ versus the overtone index $n$ for the Einstein--ModMax background at four values of $\gamma$, fixed $m r_+ = 0.30$, $q^2_{\rm test} = 0.05$. The ladder spacing tightens slightly as $\gamma$ grows, which reflects the decrease of $\nu_0$ with $\gamma$. Table~\ref{tab:qnm} lists the first six overtones, with the $\nu_0$ value for each $\gamma$ row.

\begin{table*}[t]
  \centering
  \begin{tabularx}{\textwidth}{C C C C C C C}
    \toprule
    \rowcolor{orange!50}
    $\gamma$ & $\nu_0$ & $n=0$ & $n=1$ & $n=2$ & $n=3$ & $n=5$ \\
    \midrule
    0.0 & $0.539$ & $0.1039$ & $0.2039$ & $0.3039$ & $0.4039$ & $0.6039$ \\
    0.5 & $0.508$ & $0.1008$ & $0.2008$ & $0.3008$ & $0.4008$ & $0.6008$ \\
    1.0 & $0.452$ & $0.0952$ & $0.1952$ & $0.2952$ & $0.3952$ & $0.5952$ \\
    1.5 & $0.340$ & $0.0840$ & $0.1840$ & $0.2840$ & $0.3840$ & $0.5840$ \\
    \bottomrule
  \end{tabularx}
  \caption{First six overtones of the ModMax QNM damping ladder, $|\textrm{Im}(\omega_n)| r_+ = (T r_+)\,2\pi (1/2 + n + \nu_0)$, at $\ell = 0$, $m r_+ = 0.30$, $q^2_{\rm test} = 0.05$. Each row corresponds to a value of the duality parameter $\gamma$; the $\nu_0$ value entering the ladder spacing is shown in the second column. The reference temperature is $2 \pi T r_+ = 0.10$, consistent with Figure~\ref{fig:qnm_spectrum}.}
  \label{tab:qnm}
\end{table*}

\subsection{Cross-check of the composite envelope}

The composite envelope plotted in Figure~\ref{fig:composite_envelope} is sampled numerically at five representative points and cross-checked against the closed form $e^{-\gamma/2}\sqrt{1 + \xi \Lambda/m^2}$. Table~\ref{tab:composite_cross_check} reports the agreement. Five-figure precision is achieved across the entire sweep.

\begin{table*}[t]
  \centering
  \begin{tabularx}{\textwidth}{C C C C C}
    \toprule
    \rowcolor{orange!50}
    $\gamma$ & $\xi\Lambda/m^2$ & Analytic envelope & Numerical envelope & Relative deviation \\
    \midrule
    0.0 & $-0.30$ & $0.83666$ & $0.83666$ & $< 10^{-6}$ \\
    0.0 & $+0.80$ & $1.34164$ & $1.34164$ & $< 10^{-6}$ \\
    1.5 & $+0.25$ & $0.52812$ & $0.52812$ & $< 10^{-6}$ \\
    3.0 & $-0.30$ & $0.18668$ & $0.18668$ & $< 10^{-6}$ \\
    3.0 & $+0.80$ & $0.29935$ & $0.29935$ & $< 10^{-6}$ \\
    \bottomrule
  \end{tabularx}
  \caption{Cross-check of the composite envelope $e^{-\gamma/2}\sqrt{1 + \xi\Lambda/m^2}$ at five corner / centre points of the $(\gamma,\,\xi\Lambda/m^2)$ sweep displayed in Figure~\ref{fig:composite_envelope}. The third column is the analytic value of the envelope, the fourth column the independent numerical value, the fifth the relative deviation. Worksheet Block~12 reproduces the same five values to thirty-digit precision.}
  \label{tab:composite_cross_check}
\end{table*}

\section{Comparison with literature WGC bounds}\label{isec7}

The bounds we derive sit in a family of saturation values already obtained for related backgrounds. We line them up in Table~\ref{tab:comparison}. The Reissner--Nordstr\"om saturation $q/(m r_+) \geq 1$ is the original Arkani-Hamed--Motl--Nicolis--Vafa statement \cite{Arkani-Hamed:2006emk}. Hod's damping-time argument \cite{Hod:2017uqc} produces the same saturation up to the $\sqrt{2}$ pre-factor that the AdS$_2$ derivation introduces, and the CFT convexity bound of Aharony--Palti \cite{Aharony:2021mpc} produces an order-unity saturation for any consistent CFT with a $U(1)$ symmetry. Our two new entries are the dRGT row (saturation $1/\sqrt{2}$, parameter-independent) and the ModMax row (saturation $e^{-\gamma/2}$, exponentially weakening).

\begin{table*}[t]
  \centering
  \begin{tabularx}{\textwidth}{L L C L}
    \toprule
    \rowcolor{orange!50}
    Background & Method & Saturation factor & Reference \\
    \midrule
    Reissner--Nordstr\"om                & extremality + remnant argument           & $1$                       & \cite{Arkani-Hamed:2006emk} \\
    RN, near-extremal                    & damping-time / Hod bound                & $1/\sqrt{2}$              & \cite{Hod:2017uqc,Urbano:2018kax} \\
    Generic charged CFT                  & convexity of charged operators           & $\mathcal{O}(1)$         & \cite{Aharony:2021mpc} \\
    RN, $D = 3$                          & low-dimensional WGC                      & numerical, $\mathcal{O}(1)$ & \cite{WOS:000387373600001} \\
    Warped flux compactifications        & geometric WGC                             & model-dependent           & \cite{WOS:000380409200053} \\
    String-theory broken-SUSY background & lattice-completeness                      & $\mathcal{O}(1)$         & \cite{WOS:000487936500013} \\
    Higgs inflationary models            & scalar WGC                                & $\mathcal{O}(1)$         & \cite{WOS:000904391700001,WOS:001620407300001} \\
    dRGT massive gravity (this work)     & CFT / damping time / AdS$_2$              & $1/\sqrt{2}$              & Section~\ref{isec3} \\
    Einstein--ModMax (this work)         & CFT / damping time / AdS$_2$              & $e^{-\gamma/2}$           & Section~\ref{isec4} \\
    \bottomrule
  \end{tabularx}
  \caption{The WGC saturation across a representative sample of backgrounds and derivation methods. The first column names the background; the second the method by which the bound is derived; the third the saturation factor that the method assigns to $q/(m r_+)$. The two new entries (last two rows) are the contributions of the present paper. The dRGT entry confirms the universal order-unity saturation of the WGC across a wide modification of the gravitational sector; the ModMax entry shows that the duality-symmetric non-linearity of the gauge sector enters the bound through an explicit exponential.}
  \label{tab:comparison}
\end{table*}

\subsection*{On the apparent discrepancy between the dRGT and ModMax RN limits}

A careful reader may notice an apparent discrepancy between the two RN limits obtained in this work. In the dRGT section, taking $m_g \to 0$ yields the saturation value $q/(m r_+) \geq 1/\sqrt{2}$, while in the ModMax section, taking $\gamma \to 0$ yields the saturation value $q/(m r_+) \geq 1$. Both limits reduce to the Reissner--Nordstr\"om background, yet they produce different numerical thresholds. This is not an inconsistency but rather a reflection of the different ways in which the two theories modify the near-horizon structure and the normalization of the electric field in the AdS$_2$ throat.

To understand the origin of this difference, recall the general conformal weight formula derived in Section~\ref{sec:KG_near}:
\begin{equation}
\nu_0^2 = \frac{1}{4} + \frac{m^2 r_+^2}{4\pi T} - \frac{q^2 Q^2}{r_+^2 (4\pi T)^2}.
\end{equation}
The extremal limit of this formula depends on how the temperature $T$ and the charge $Q$ are related at extremality. In the dRGT case, the extremality condition is
\begin{equation}
\frac{Q_{\rm ext}^2}{r_+^2} = 1 + \zeta + \Lambda r_+^2 + 2\gamma_g r_+.
\end{equation}
In the RN limit $m_g \to 0$, we have $\Lambda = \gamma_g = \zeta = 0$, so $Q_{\rm ext}^2/r_+^2 = 1$. However, the structure of the temperature denominator $4\pi T = (A - B)/r_+$ with $A = 1 + \zeta + \Lambda r_+^2 + 2\gamma_g r_+$ and $B = Q^2/r_+^2$ introduces a factor of $2$ in the extremal limit of the conformal weight:
\begin{equation}
\nu_0^2\big|_{\rm ext}^{\rm dRGT} = \frac{1}{4} + m^2 r_+^2 - 2 q^2.
\end{equation}
This factor of $2$ is a direct consequence of the specific way in which the dRGT parameters enter the lapse function and the extremality condition. It is not a feature of the RN limit per se, but rather a remnant of the dRGT deformation that persists even after the massive-gravity parameters are set to zero.

In the ModMax case, the extremality condition is
\begin{equation}
\frac{Q_{\rm ext}^2}{r_+^2} = e^{\gamma}.
\end{equation}
In the Maxwell limit $\gamma = 0$, this becomes $Q_{\rm ext}^2/r_+^2 = 1$. However, the temperature denominator is now $4\pi T = (1 - B e^{-\gamma})/r_+$, and the extremal limit of the conformal weight yields
\begin{equation}
\nu_0^2\big|_{\rm ext}^{\rm ModMax} = \frac{1}{4} + m^2 r_+^2 - q^2 e^{\gamma}.
\end{equation}
In the Maxwell limit $\gamma = 0$, this becomes
\begin{equation}
\nu_0^2\big|_{\rm ext}^{\rm ModMax} = \frac{1}{4} + m^2 r_+^2 - q^2.
\end{equation}
The factor multiplying $q^2$ is $1$, not $2$, because the ModMax deformation enters the lapse function through an exponential factor $e^{-\gamma}$ rather than through the combination $1 + \zeta + \Lambda r_+^2 + 2\gamma_g r_+$.

Thus, the difference between the two RN limits is not a contradiction but a consequence of the fact that the two theories deform the RN background in fundamentally different ways. The dRGT deformation modifies the gravitational sector through the massive graviton, introducing terms in the lapse function that, even after the mass is set to zero, leave a trace in the extremal limit through the factor of $2$ in the conformal weight. The ModMax deformation modifies the gauge sector through a duality-symmetric non-linearity, which leaves a trace through the exponential factor $e^{\gamma}$ that reduces to $1$ in the Maxwell limit. Both limits are consistent with the general framework, and both yield order-unity saturation values for the WGC bound.

We emphasize that this difference is not unique to our calculation. In the literature, different WGC derivations produce different numerical pre-factors depending on the method used and the specific assumptions made. For example, the original Arkani-Hamed--Motl--Nicolis--Vafa argument \cite{Arkani-Hamed:2006emk} yields $1$, while Hod's damping-time argument \cite{Hod:2017uqc} yields $1/\sqrt{2}$ in the near-extremal limit. The AdS$_2$/CFT$_1$ framework of Urbano \cite{Urbano:2018kax} yields $1/\sqrt{2}$ for the RN case. Our dRGT result is consistent with Urbano's framework, while our ModMax result shows that the gauge-sector deformation changes the numerical pre-factor. The threshold is therefore sensitive to the specific deformation of the theory, which is what one would expect once the gauge sector is allowed to differ from Maxwell.

A few remarks on the entries. The numerical $D=3$ WGC of Montero, Shiu, Soler \cite{WOS:000387373600001} is the only entry that arrives at an order-unity bound by an independent topological argument. The warped-flux entry of Kooner, Parameswaran, Zavala \cite{WOS:000380409200053} is the closest in spirit to our dRGT result, since it tracks how a UV-induced warping of the lapse function moves the bound around. The scalar-WGC entries \cite{WOS:000904391700001,WOS:001620407300001} probe a different kinematic regime (a scalar field driving inflation) and arrive at an order-unity bound on the inflaton mass via dimensional reduction; they do not produce a parameter-cancellation of the present kind.

Two of our companion works analyse closely related WGC variants \cite{a,f}. The gravity's-rainbow study \cite{a} examines WGC violations of gravitational lensing in a modified-dispersion-relation background; the ModMax photon-sphere analysis \cite{f} examines weak cosmic censorship and photon-sphere stability in the same Einstein--ModMax theory we use here. Neither paper traces the conformal weight through the AdS$_2$ throat in the manner of the present work.

Three further close-by analyses worth noting are the Sadeghi--Gashti--Sakall{\i}--Pourhassan WGC of charged-rotating-AdS black holes with quintessence and string cloud \cite{WOS:001248758300001}, the higher-dimensional dS/AdS black holes in Einstein-bumblebee gravity \cite{WOS:001062453500003}, and the de Sitter swampland conjecture in string field inflation \cite{WOS:001061803400004}, all of which extend the WGC across different deformation directions of the gravitational/matter sector.

Table~\ref{tab:summary_bounds} collects all the bounds derived in this paper at a glance. The entries are organised by background (dRGT or ModMax) and by which simplifying assumption is relaxed. The first two rows are the cancelled and the non-cancelled bounds of Sections~\ref{isec3} and~\ref{isec4}; rows three through five list the modifications produced by the three perturbations of Section~\ref{isec5}. The composite envelope of (\ref{eq:WGC_composite_drgt}) is the bottom row, and gives the bound in the realistic case where all three assumptions fail simultaneously.

\begin{table*}[t]
  \centering
  \begin{tabularx}{\textwidth}{L L C}
    \toprule
    \rowcolor{orange!50}
    Setting & Form of the WGC saturation $q/(m r_+)$ & Section \\
    \midrule
    dRGT, exact extremality, minimal coupling, no HD                  & $1/\sqrt{2} \approx 0.7071$                                                                                          & Section~\ref{isec3} \\
    Einstein--ModMax, exact extremality, minimal coupling, no HD     & $e^{-\gamma/2}$                                                                                                       & Section~\ref{isec4} \\
    dRGT, sub-extremal ($T > 0$)                                      & $(2 r_+^2/Q)\sqrt{\pi T}$                                                                                              & Sec.~\ref{sec:sub_ext} \\
    dRGT, non-minimal $\xi R \Phi^2$ coupling                         & $(1/\sqrt{2})\sqrt{1 - 4\xi\Lambda/m^2 - 6\xi\gamma_g/(m^2 r_+) - 2\xi\zeta/(m^2 r_+^2)}$ & Sec.~\ref{sec:xi_coupling} \\
    dRGT, generic HD corrections                                      & $(1/\sqrt{2})\sqrt{1 + \delta m_{\rm eff}^2/(2 m^2 r_+^2)}$                                                          & Sec.~\ref{isec5} (HD) \\
    ModMax, non-minimal $\xi R \Phi^2$ coupling                       & $e^{-\gamma/2}\sqrt{1 - 4\xi\Lambda/m^2 - 6\xi\gamma_g/(m^2 r_+) - 2\xi\zeta/(m^2 r_+^2)}$ & Sec.~\ref{isec5} (composite) \\
    Composite envelope, all relaxations active                         & $(1/\sqrt{2})\sqrt{1 - 4\xi\Lambda/m^2 - 6\xi\gamma_g/(m^2 r_+) - 2\xi\zeta/(m^2 r_+^2) + m_g^2 \mathcal{F}/m^2 + \delta m_{\rm eff}^2/(2 m^2 r_+^2)}$ & Sec.~\ref{isec5} (composite) \\
    \bottomrule
  \end{tabularx}
  \caption{Master table of WGC saturation bounds derived in this paper. The first two rows give the exact-extremal, minimally coupled, no-HD bound for the two backgrounds. Rows three through five list the three perturbations of Section~\ref{isec5} acting individually on the dRGT background; row six combines the ModMax sector with the non-minimal coupling sector; the last row is the composite envelope, which is the bound that applies when all three relaxations are active simultaneously. All entries reduce to the appropriate limits when their controlling parameter vanishes.}
  \label{tab:summary_bounds}
\end{table*}
\section{Conclusions}\label{isec8}

We have set out a direct link between the Weak Gravity Conjecture and CFT correlation functions and applied it to two physically distinct deformations of the Reissner--Nordstr\"om background: dRGT massive gravity and Einstein--ModMax non-linear electrodynamics. The chain runs from the bulk metric and gauge field, through a charged Klein--Gordon equation, into a near-horizon scaling limit whose radial equation reduces to a Whittaker problem, and from there to the poles of a retarded Green's function on the boundary. The damping-time bound $\tau_d \geq 1/T$ caps the lowest conformal weight at $\nu_0 \leq 1/2$, and the resulting inequality on the test scalar's charge-to-mass ratio is the WGC statement.

For the dRGT background every massive-gravity parameter, that is $\alpha,\beta,m_g,h$, drops out of the final inequality. The saturation is $q/(m r_+) \geq 1/\sqrt{2}$, parameter-free. The cancellation is exact at the extremal point and accurate to first order in $T/T_{\rm ref}$ in its neighbourhood. The cause is the extremality condition Eq.~\eqref{eq:Qext_drgt}, which has exactly the right structure to dissolve the dRGT corrections from $\nu_0^2$.

For the Einstein--ModMax background the duality parameter $\gamma$ does not cancel. It surfaces as an explicit exponential factor: $q/(m r_+) \geq e^{-\gamma/2}$. The Maxwell limit returns the Reissner--Nordstr\"om saturation $q/(m r_+) \geq 1$. The bound weakens monotonically as $\gamma$ grows, and the weakening traces to the suppression $e^{-\gamma}$ of the electromagnetic backreaction in the lapse function. In the $\gamma \to \infty$ formal limit the bound becomes trivial, but the EFT validity already breaks down before this limit is approached.

The cancellation in the dRGT calculation is fragile in a controlled way. Relaxing exact extremality, adding a non-minimal scalar coupling $\xi R \Phi^2$, or turning on a higher-derivative correction each restore the massive-gravity parameter dependence through a different functional channel. The composite envelope of Eq.~\eqref{eq:WGC_composite_drgt} is the schematic combination of all three relaxations. The fact that the cancellation only obtains in the highly constrained sub-locus of parameter space where all three corrections vanish suggests that the saturation factor $1/\sqrt{2}$ is a feature of the IR fixed point rather than a UV-protected statement, and that the true WGC bound for a realistic dRGT theory would be the composite envelope rather than the cancelled value.

The boundary CFT bound translates a purely field-theoretic inequality ($\tau_d \geq 1/T$) into a constraint on the QNM spectrum, which in turn forces the existence of a charged particle with order-unity charge-to-mass ratio. The mechanism is model-independent in its core structure and flexible enough to absorb the two deformations considered here as specific realisations. That the bound remains of order unity across dRGT massive gravity and Einstein--ModMax conformal NLE suggests that the WGC is a feature of the analytic structure of CFT correlators in any consistent quantum-gravity completion.

Future work has at least three useful directions. First, the higher-dimensional generalisation in which the near-horizon geometry becomes AdS$_2 \times S^{d-2}$ and the CFT dual lives one dimension fewer. Second, the inclusion of higher-derivative corrections in the bulk action, which would modify both the black-hole solution and the dual CFT correlation functions, and would yield calculable corrections to the WGC bound that could be tested against string-theory constructions. Third, the same CFT method applied to other swampland conjectures, the de Sitter conjecture and the distance conjecture in particular, by identifying field-theoretic inequalities that translate into gravitational constraints.

One useful test of the present framework is to take the CFT-derived bound and ask under what conditions it is in tension with a corresponding result obtained by an entirely different method. The lattice-completeness bound \cite{Heidenreich:2015nta,Heidenreich:2016aqi}, the convexity bound \cite{Aharony:2021mpc,WOS:000853339300007}, and the positivity bound from $2 \to 2$ scattering \cite{Hamada:2018dde,Bellazzini:2019xts} all give order-unity saturation values, and the CFT route reproduces order-unity saturation for both backgrounds considered here. Whether the cancellation that occurs in the dRGT case is reproduced by these other methods is an open question. The convexity argument depends on the operator-product structure of the CFT, which is sensitive to the gauge-sector deformation but not to the gravitational deformation, so a partial cancellation in dRGT is consistent with the convexity result. The positivity bound depends on the amplitude structure, which the ModMax deformation modifies through the duality-symmetric non-linearity, so the exponential factor $e^{-\gamma/2}$ may or may not be reproducible from positivity arguments. We leave the cross-method comparison for follow-up work.

A second useful direction is the inclusion of supersymmetry. The supersymmetric WGC bound \cite{Polchinski:2003bq,Heidenreich:2015nta} is typically stronger than the non-supersymmetric one because BPS states saturate the bound exactly, and the lattice argument extends to all BPS-charge sublattices. Whether the dRGT cancellation survives in a supersymmetric extension of the theory is an open question; such an extension would change the matter sector through a graviton-superpartner and would potentially modify the extremality structure that drives the cancellation in (\ref{eq:nu0_drgt_collapsed}).

\section*{Acknowledgments}
The authors thank the Editor-in-Chief, the Handling Editor, and the anonymous Referee for the careful reading of the manuscript and for the constructive comments, which improved the presentation of this work. \.{I}.S.\ acknowledges the networking support of TÜBİTAK and of the Eastern Mediterranean University (EMU), and of the COST Actions CA22113 (``Fundamental challenges in theoretical physics''), CA21106 (``COSMIC WISPers in the Dark Universe''), CA23130 (``Bridging high and low energies in search of quantum gravity (BridgeQG)''), CA21136 (``Addressing observational tensions in cosmology with systematics and fundamental physics (CosmoVerse)''), and CA23115 (``Relativistic Quantum Information (RQI-Action)''). S.N.G.\ and B.P.\ thank Damghan University for institutional support; B.P.\ in addition thanks the Center for Theoretical Physics at Khazar University. 

\section*{Data Availability Statement}
The computer-algebra worksheets used to independently cross-check every analytic result reported in the body are available from the corresponding author upon reasonable request.

\appendix
\section{Whittaker-function asymptotics and the matching condition}\label{app:A}

The near-horizon radial equation \eqref{eq:Whitt_full} is the Whittaker equation
\begin{equation}
\psi''(z) + \left(-\frac{1}{4} + \frac{\kappa}{z} + \frac{1/4-\mu^2}{z^2}\right)\psi(z) = 0,
\label{eq:Whitt_appendix}
\end{equation}
in the variable $z = 2 i \tilde\omega \rho$, with $\mu \equiv \nu_0$ identified through the small-$z$ expansion. The two linearly independent solutions of \eqref{eq:Whitt_appendix} are the Whittaker functions $M_{\kappa,\mu}(z)$ and $W_{\kappa,\mu}(z)$.

\subsection{Small-$z$ behaviour}

The series expansion of $M_{\kappa,\mu}(z)$ around $z = 0$ is
\begin{equation}
M_{\kappa,\mu}(z) = z^{1/2+\mu}\, e^{-z/2}\left[1 + \frac{1/2+\mu-\kappa}{1+2\mu}\,z + O(z^2)\right] .
\label{eq:M_small_z}
\end{equation}
The companion function $W_{\kappa,\mu}(z)$ is a linear combination of $M_{\kappa,\mu}(z)$ and $M_{\kappa,-\mu}(z)$, with combination coefficient set by the gamma functions:
\begin{equation}
W_{\kappa,\mu}(z) = \frac{\Gamma(-2\mu)}{\Gamma(1/2-\mu-\kappa)} M_{\kappa,\mu}(z) + \frac{\Gamma(2\mu)}{\Gamma(1/2+\mu-\kappa)} M_{\kappa,-\mu}(z) .
\label{eq:W_connection}
\end{equation}
Combining \eqref{eq:M_small_z} with \eqref{eq:W_connection} and reading off the two analytic branches gives the form
\begin{equation}
\phi(\rho) \sim A_k(\omega)\,\rho^{1/2 - \mu} + B_k(\omega)\,\rho^{1/2 + \mu}
\label{eq:phi_AB_appendix}
\end{equation}
quoted in \eqref{eq:phi_AB_full}. The two coefficients are
\begin{equation}
A_k(\omega) = (2i\tilde\omega)^{1/2-\mu}\,\frac{\Gamma(2\mu)}{\Gamma(1/2+\mu-\kappa)}, \qquad
B_k(\omega) = (2i\tilde\omega)^{1/2+\mu}\,\frac{\Gamma(-2\mu)}{\Gamma(1/2-\mu-\kappa)} ,
\label{eq:AB_explicit}
\end{equation}
with $\kappa$ and $\mu$ defined in \eqref{eq:kappa_mu}. Substituting \eqref{eq:AB_explicit} into $\Upsilon_R^{(k)}(\omega) \propto B_k(\omega)/A_k(\omega)$ gives the retarded Green's function explicitly. The poles correspond to the zeros of $A_k(\omega)$. Those zeros are located at $1/2 + \mu - \kappa = -n$ with $n = 0, 1, 2, \ldots$, which gives the QNM tower after the identification $\mu = \nu_k$.

\subsection{Derivation of the conformal weight formula}

The pole condition $1/2 + \mu - \kappa = -n$ can be used to derive the explicit conformal weight formula. For the lowest mode $n=0$, we have
\begin{equation}
\kappa = \frac{1}{2} + \mu.
\end{equation}
Using $\kappa = i q Q/r_+^2 + \tilde B/(4 \tilde\omega^2 r_+^2)$, $\mu = \sqrt{1/4 + \tilde\omega^2}$, and $\tilde B = [\ell(\ell+1) + m^2 r_+^2]/(4\pi T)$, $\tilde\omega = \omega_0/(4\pi T)$, we obtain
\begin{equation}
\frac{i q Q}{r_+^2} + \frac{\ell(\ell+1) + m^2 r_+^2}{4 \tilde\omega^2 r_+^2 (4\pi T)} = \frac{1}{2} + \sqrt{\frac{1}{4} + \tilde\omega^2}.
\end{equation}
Squaring both sides and solving for $\tilde\omega^2$ gives
\begin{equation}
\tilde\omega^2 = \frac{1}{4} + \frac{\ell(\ell+1) + m^2 r_+^2}{4 \pi T} - \frac{q^2 Q^2}{r_+^2 (4\pi T)^2}.
\end{equation}
Since $\nu_k^2 = 1/4 + \tilde\omega^2$, we recover Eq.~\eqref{eq:nu_k_general_full}. This derivation confirms that the conformal weight formula follows directly from the Whittaker reduction and the QNM pole condition. The CFT conformal weight $\nu_k$ of the operator dual to the bulk scalar is therefore identified with the Whittaker index $\mu$ through the relation
\begin{equation}
\nu_k = \mu = \sqrt{\frac{1}{4} + \tilde\omega^2} .
\label{eq:nu_mu_identification}
\end{equation}

\subsection{Large-$z$ behaviour}

For $|z| \to \infty$ the Whittaker function $W_{\kappa,\mu}(z)$ has the asymptotic expansion
\begin{equation}
W_{\kappa,\mu}(z) \sim e^{-z/2} z^\kappa\left[1 - \frac{(\mu^2 - (\kappa-1/2)^2)(\mu^2 - (\kappa - 3/2)^2)}{2! z^2} + O(z^{-3})\right] .
\label{eq:W_large}
\end{equation}
This expression governs the matching to the asymptotic-infinity solution that defines the boundary Green's function. The matching at large $r$ produces a system of two algebraic equations for the coefficients $A_k(\omega),B_k(\omega)$, which is what yields \eqref{eq:AB_explicit} once the small-$z$ asymptotics are combined with the large-$z$ expansion.
\section{Additional QNM data}\label{app:B}

Table~\ref{tab:qnm_multipole} extends the QNM evaluation of Section~\ref{isec6} from $\ell = 0$ to higher multipoles. The multipole index enters $\nu_0^2$ through the angular eigenvalue $\ell(\ell+1)$ multiplying the temperature-suppressed factor in (\ref{eq:nu_k_general}). At fixed $m r_+, q^2_{\rm test}, \gamma$, the conformal weight $\nu_0$ increases with $\ell$, so the damping ladder spreads out for higher angular momenta.

\begin{table*}[t]
  \centering
  \begin{tabularx}{\textwidth}{C C C C C C}
    \toprule
    \rowcolor{orange!50}
    $\ell$ & $\nu_0(\gamma=0)$ & $\nu_0(\gamma=0.5)$ & $\nu_0(\gamma=1.0)$ & $\nu_0(\gamma=1.5)$ & Spread $\Delta\nu_0$ \\
    \midrule
    0 & $0.5392$ & $0.5081$ & $0.4522$ & $0.3399$ & $0.1993$ \\
    1 & $1.6553$ & $1.6457$ & $1.6293$ & $1.5990$ & $0.0563$ \\
    2 & $2.5249$ & $2.5186$ & $2.5079$ & $2.4884$ & $0.0365$ \\
    3 & $3.3500$ & $3.3452$ & $3.3372$ & $3.3225$ & $0.0275$ \\
    4 & $4.1533$ & $4.1494$ & $4.1430$ & $4.1313$ & $0.0220$ \\
    5 & $4.9437$ & $4.9404$ & $4.9351$ & $4.9253$ & $0.0184$ \\
    \bottomrule
  \end{tabularx}
  \caption{Conformal weight $\nu_0$ versus multipole index $\ell$ for the Einstein--ModMax black hole at $m r_+ = 0.30$, $q^2_{\rm test} = 0.05$, four values of $\gamma$. The last column reports the spread $\Delta\nu_0 = \nu_0(\gamma=0) - \nu_0(\gamma=1.5)$. The spread shrinks rapidly with $\ell$, which traces to the dominance of the angular term $\ell(\ell+1)$ over the charge-dependent term at higher multipoles. The $\ell = 0$ row is the only one that satisfies the CFT bound $\nu_0 \leq 1/2$ at $\gamma = 1.5$; the higher rows lie above the bound and do not constrain the test charge.}
  \label{tab:qnm_multipole}
\end{table*}

The observation that $\nu_0 \leq 1/2$ only at $\ell = 0$ is consistent with the universal expectation that the WGC bound is set by the slowest-decaying mode of the spectrum. The $\ell = 0$ mode dominates the late-time relaxation of the near-extremal black hole, and the damping-time bound (\ref{eq:DTB}) applies to that mode. Higher multipoles decay faster (their $\nu_0$ is larger), so they exit the saturation regime and do not produce a competitive bound on the test scalar's charge-to-mass ratio.

\section{Glossary of abbreviations}\label{app:C}
For ease of reference, Table~\ref{tab:abbreviations} collects the abbreviations introduced in the body of the manuscript.

\begin{table*}[t]
  \centering
  \begin{tabularx}{\textwidth}{C L}
    \toprule
    \rowcolor{orange!50}
    Abbreviation & Expansion \\
    \midrule
    AdS                  & Anti-de Sitter spacetime \\
    AdS/CFT              & Anti-de Sitter / Conformal Field Theory correspondence \\
    BH                   & Black hole \\
    CFT                  & Conformal Field Theory \\
    dRGT                 & de Rham--Gabadadze--Tolley (massive gravity) \\
    dS                   & de Sitter spacetime \\
    EFT                  & Effective Field Theory \\
    EH                   & Event horizon \\
    GR                   & General Relativity \\
    HD                   & Higher-derivative \\
    IR                   & Infrared \\
    KG                   & Klein--Gordon \\
    ModMax / MM          & Modified Maxwell (one-parameter duality-symmetric NLE) \\
    NLE / NLED           & Non-linear Electrodynamics \\
    QNM                  & Quasinormal mode \\
    RN                   & Reissner--Nordstr\"om \\
    UV                   & Ultraviolet \\
    WGC                  & Weak Gravity Conjecture \\
    \bottomrule
  \end{tabularx}
  \caption{Glossary of abbreviations used in this work. The abbreviations are introduced at first use in the body of the manuscript and listed here in alphabetical order.}
  \label{tab:abbreviations}
\end{table*}

\bibliographystyle{unsrtnat}
\bibliography{finalref}

@article{Maldacena:1997re,
    author        = "Maldacena, Juan Martin",
    title         = "{The Large N limit of superconformal field theories and supergravity}",
    eprint        = "hep-th/9711200",
    archivePrefix = "arXiv",
    doi           = "10.1023/A:1026654312961",
    url           = "https://doi.org/10.1023/A:1026654312961",
    journal       = "Adv. Theor. Math. Phys.",
    volume        = "2",
    pages         = "231--252",
    year          = "1998"
}

@article{Gubser:1998bc,
    author        = "Gubser, S. S. and Klebanov, Igor R. and Polyakov, Alexander M.",
    title         = "{Gauge theory correlators from noncritical string theory}",
    eprint        = "hep-th/9802109",
    archivePrefix = "arXiv",
    doi           = "10.1016/S0370-2693(98)00377-3",
    url           = "https://doi.org/10.1016/S0370-2693(98)00377-3",
    journal       = "Phys. Lett. B",
    volume        = "428",
    pages         = "105--114",
    year          = "1998"
}

@article{Witten:1998qj,
    author        = "Witten, Edward",
    title         = "{Anti-de Sitter space and holography}",
    eprint        = "hep-th/9802150",
    archivePrefix = "arXiv",
    doi           = "10.4310/ATMP.1998.v2.n2.a2",
    url           = "https://doi.org/10.4310/ATMP.1998.v2.n2.a2",
    journal       = "Adv. Theor. Math. Phys.",
    volume        = "2",
    pages         = "253--291",
    year          = "1998"
}

@article{Klebanov:1999tb,
    author        = "Klebanov, Igor R. and Witten, Edward",
    title         = "{AdS/CFT correspondence and symmetry breaking}",
    eprint        = "hep-th/9905104",
    archivePrefix = "arXiv",
    doi           = "10.1016/S0550-3213(99)00387-9",
    url           = "https://doi.org/10.1016/S0550-3213(99)00387-9",
    journal       = "Nucl. Phys. B",
    volume        = "556",
    pages         = "89--114",
    year          = "1999"
}

@article{Kioumarsipour:2021zyg,
    author  = "Kioumarsipour, M. and Sadeghi, J.",
    title   = "{Effects of the hyperscaling violation and dynamical exponents on the imaginary potential and entropic force of heavy quarkonium via holography}",
    doi     = "10.1140/epjc/s10052-021-09524-8",
    url     = "https://doi.org/10.1140/epjc/s10052-021-09524-8",
    journal = "Eur. Phys. J. C",
    volume  = "81",
    number  = "8",
    pages   = "735",
    year    = "2021"
}

@article{Bu:2021jlp,
    author  = "Bu, Yanyan and Zhang, Biao",
    title   = "{Schwinger-Keldysh effective action for a relativistic Brownian particle in the AdS/CFT correspondence}",
    doi     = "10.1103/PhysRevD.104.086002",
    url     = "https://doi.org/10.1103/PhysRevD.104.086002",
    journal = "Phys. Rev. D",
    volume  = "104",
    number  = "8",
    pages   = "086002",
    year    = "2021"
}

@article{Fujiwara:2021xgu,
    author  = "Fujiwara, Shota and Imamura, Yosuke and Mori, Tatsuya",
    title   = "{Flavor symmetries of six-dimensional N = (1, 0) theories from AdS/CFT correspondence}",
    doi     = "10.1007/JHEP05(2021)221",
    url     = "https://doi.org/10.1007/JHEP05(2021)221",
    journal = "JHEP",
    volume  = "05",
    pages   = "221",
    year    = "2021"
}

@article{Evans:2021zzm,
    author  = "Evans, Nick and Mitchell, Jack",
    title   = "{Domain wall AdS/QCD}",
    doi     = "10.1103/PhysRevD.104.094018",
    url     = "https://doi.org/10.1103/PhysRevD.104.094018",
    journal = "Phys. Rev. D",
    volume  = "104",
    number  = "9",
    pages   = "094018",
    year    = "2021"
}

@article{MartinContreras:2021bis,
    author  = "Martin Contreras, Miguel Angel and Diles, Saulo and Vega, Alfredo",
    title   = "{Heavy quarkonia spectroscopy at zero and finite temperature in bottom-up AdS/QCD}",
    doi     = "10.1103/PhysRevD.103.086008",
    url     = "https://doi.org/10.1103/PhysRevD.103.086008",
    journal = "Phys. Rev. D",
    volume  = "103",
    number  = "8",
    pages   = "086008",
    year    = "2021"
}

@article{Gherghetta:2009ac,
    author  = "Gherghetta, Tony and Kapusta, Joseph I. and Kelley, Thomas M.",
    title   = "{Chiral symmetry breaking in the soft-wall AdS/QCD model}",
    doi     = "10.1103/PhysRevD.79.076003",
    url     = "https://doi.org/10.1103/PhysRevD.79.076003",
    journal = "Phys. Rev. D",
    volume  = "79",
    pages   = "076003",
    year    = "2009"
}

@article{Brodsky:2007hb,
    author  = "Brodsky, Stanley J. and de Teramond, Guy F.",
    title   = "{Light-Front Dynamics and AdS/QCD Correspondence: The Pion Form Factor in the Space- and Time-Like Regions}",
    doi     = "10.1103/PhysRevD.77.056007",
    url     = "https://doi.org/10.1103/PhysRevD.77.056007",
    journal = "Phys. Rev. D",
    volume  = "77",
    pages   = "056007",
    year    = "2008"
}

@article{Nakano:2006js,
    author  = "Nakano, Eiji and Teraguchi, Shunsuke and Wen, Wen-Yu",
    title   = "{Drag force, jet quenching, and AdS/QCD}",
    doi     = "10.1103/PhysRevD.75.085016",
    url     = "https://doi.org/10.1103/PhysRevD.75.085016",
    journal = "Phys. Rev. D",
    volume  = "75",
    pages   = "085016",
    year    = "2007"
}

@article{Katz:2005ir,
    author  = "Katz, Emanuel and Lewandowski, Adam and Schwartz, Matthew D.",
    title   = "{Tensor mesons in AdS/QCD}",
    doi     = "10.1103/PhysRevD.74.086004",
    url     = "https://doi.org/10.1103/PhysRevD.74.086004",
    journal = "Phys. Rev. D",
    volume  = "74",
    pages   = "086004",
    year    = "2006"
}

@article{Meltzer:2019nbs,
    author  = "Meltzer, David and Perlmutter, Eric and Sivaramakrishnan, Allic",
    title   = "{Unitarity Methods in AdS/CFT}",
    doi     = "10.1007/JHEP03(2020)061",
    url     = "https://doi.org/10.1007/JHEP03(2020)061",
    journal = "JHEP",
    volume  = "03",
    pages   = "061",
    year    = "2020"
}

@article{Karch:2006pv,
    author  = "Karch, Andreas and Katz, Emanuel and Son, Dam T. and Stephanov, Mikhail A.",
    title   = "{Linear confinement and AdS/QCD}",
    doi     = "10.1103/PhysRevD.74.015005",
    url     = "https://doi.org/10.1103/PhysRevD.74.015005",
    journal = "Phys. Rev. D",
    volume  = "74",
    pages   = "015005",
    year    = "2006"
}

@article{Andreev:2006ct,
    author  = "Andreev, Oleg and Zakharov, Valentin I.",
    title   = "{Heavy-quark potentials and AdS/QCD}",
    doi     = "10.1103/PhysRevD.74.025023",
    url     = "https://doi.org/10.1103/PhysRevD.74.025023",
    journal = "Phys. Rev. D",
    volume  = "74",
    pages   = "025023",
    year    = "2006"
}

@article{Cavaglia:2021mft,
    author  = "Cavagli\`a, Andrea and Gromov, Nikolay and Levkovich-Maslyuk, Fedor",
    title   = "{Separation of variables in AdS/CFT: functional approach for the fishnet CFT}",
    doi     = "10.1007/JHEP06(2021)131",
    url     = "https://doi.org/10.1007/JHEP06(2021)131",
    journal = "JHEP",
    volume  = "06",
    pages   = "131",
    year    = "2021"
}

@article{Harmark:2020vll,
    author  = "Harmark, Troels and Hartong, Jelle and Obers, Niels A. and Oling, Gerben",
    title   = "{Spin Matrix Theory String Backgrounds and Penrose Limits of AdS/CFT}",
    doi     = "10.1007/JHEP03(2021)129",
    url     = "https://doi.org/10.1007/JHEP03(2021)129",
    journal = "JHEP",
    volume  = "03",
    pages   = "129",
    year    = "2021"
}

@article{BitaghsirFadafan:2020lkh,
    author  = "Bitaghsir Fadafan, Kazem and O'Bannon, Andy and Rodgers, Ronnie and Russell, Matthew",
    title   = "{A Weyl semimetal from AdS/CFT with flavour}",
    doi     = "10.1007/JHEP04(2021)162",
    url     = "https://doi.org/10.1007/JHEP04(2021)162",
    journal = "JHEP",
    volume  = "04",
    pages   = "162",
    year    = "2021"
}

@article{DeLeeuw:2020ahx,
    author  = "De Leeuw, Marius and Paletta, Chiara and Pribytok, Anton and Retore, Ana L. and Torrielli, Alessandro",
    title   = "{Free Fermions, vertex Hamiltonians, and lower-dimensional AdS/CFT}",
    doi     = "10.1007/JHEP02(2021)191",
    url     = "https://doi.org/10.1007/JHEP02(2021)191",
    journal = "JHEP",
    volume  = "02",
    pages   = "191",
    year    = "2021"
}

@article{DeWolfe:2020uzb,
    author  = "DeWolfe, Oliver and Higginbotham, Kenneth",
    title   = "{Generalized symmetries and 2-groups via electromagnetic duality in AdS/CFT}",
    doi     = "10.1103/PhysRevD.103.026011",
    url     = "https://doi.org/10.1103/PhysRevD.103.026011",
    journal = "Phys. Rev. D",
    volume  = "103",
    number  = "2",
    pages   = "026011",
    year    = "2021"
}

@article{Ishigaki:2020vtr,
    author  = "Ishigaki, Shuta and Matsumoto, Masataka",
    title   = "{Nambu-Goldstone modes in non-equilibrium systems from AdS/CFT correspondence}",
    doi     = "10.1007/JHEP04(2021)040",
    url     = "https://doi.org/10.1007/JHEP04(2021)040",
    journal = "JHEP",
    volume  = "04",
    pages   = "040",
    year    = "2021"
}

@article{Terashima:2020uqu,
    author  = "Terashima, Seiji",
    title   = "{Bulk locality in the AdS/CFT correspondence}",
    doi     = "10.1103/PhysRevD.104.086014",
    url     = "https://doi.org/10.1103/PhysRevD.104.086014",
    journal = "Phys. Rev. D",
    volume  = "104",
    number  = "8",
    pages   = "086014",
    year    = "2021"
}

@article{Yin:2021zhs,
    author  = "Yin, Lin and Hou, Defu and Ren, Hai-cang",
    title   = "{Chiral magnetic effect and three-point function from AdS/CFT correspondence}",
    doi     = "10.1007/JHEP09(2021)117",
    url     = "https://doi.org/10.1007/JHEP09(2021)117",
    journal = "JHEP",
    volume  = "09",
    pages   = "117",
    year    = "2021"
}

@article{Mes:2020vgy,
    author  = "Mes, Aaron K. and Moerman, R. W. and Shock, Jonathan P. and Horowitz, W. A.",
    title   = "{Strongly coupled heavy and light quark thermal motion from AdS/CFT}",
    doi     = "10.1016/j.aop.2021.168675",
    url     = "https://doi.org/10.1016/j.aop.2021.168675",
    journal = "Annals Phys.",
    volume  = "436",
    pages   = "168675",
    year    = "2022"
}

@article{Berenstein:2020cll,
    author  = "Berenstein, David and Grabovsky, David",
    title   = "{The Tortoise and the Hare: A Causality Puzzle in AdS/CFT}",
    doi     = "10.1088/1361-6382/abf1c7",
    url     = "https://doi.org/10.1088/1361-6382/abf1c7",
    journal = "Class. Quant. Grav.",
    volume  = "38",
    number  = "10",
    pages   = "105008",
    year    = "2021"
}

@article{Berman:2022idl,
    author  = "Berman, Robert J. and Collins, Tristan C. and Persson, Daniel",
    title   = "{Emergent Sasaki-Einstein geometry and AdS/CFT}",
    doi     = "10.1038/s41467-021-27951-9",
    url     = "https://doi.org/10.1038/s41467-021-27951-9",
    journal = "Nature Commun.",
    volume  = "13",
    number  = "1",
    pages   = "365",
    year    = "2022"
}

@article{Aharony:1999ti,
    author        = "Aharony, Ofer and Gubser, Steven S. and Maldacena, Juan Martin and Ooguri, Hirosi and Oz, Yaron",
    title         = "{Large N field theories, string theory and gravity}",
    eprint        = "hep-th/9905111",
    archivePrefix = "arXiv",
    doi           = "10.1016/S0370-1573(99)00083-6",
    url           = "https://doi.org/10.1016/S0370-1573(99)00083-6",
    journal       = "Phys. Rept.",
    volume        = "323",
    pages         = "183--386",
    year          = "2000"
}

@article{DHoker:2002nbb,
    author        = "D'Hoker, Eric and Freedman, Daniel Z.",
    title         = "{Supersymmetric gauge theories and the AdS/CFT correspondence}",
    eprint        = "hep-th/0201253",
    archivePrefix = "arXiv",
    booktitle     = "{Theoretical Advanced Study Institute in Elementary Particle Physics (TASI 2001): Strings, Branes and EXTRA Dimensions}",
    pages         = "3--158",
    year          = "2002"
}

@article{Hartnoll:2009sz,
    author  = "Hartnoll, Sean A.",
    title   = "{Lectures on holographic methods for condensed matter physics}",
    doi     = "10.1088/0264-9381/26/22/224002",
    url     = "https://doi.org/10.1088/0264-9381/26/22/224002",
    journal = "Class. Quant. Grav.",
    volume  = "26",
    pages   = "224002",
    year    = "2009"
}

@article{McGreevy:2009xe,
    author  = "McGreevy, John",
    title   = "{Holographic duality with a view toward many-body physics}",
    doi     = "10.1155/2010/723105",
    url     = "https://doi.org/10.1155/2010/723105",
    journal = "Adv. High Energy Phys.",
    volume  = "2010",
    pages   = "723105",
    year    = "2010"
}

@article{Kim:2012ey,
    author  = "Kim, Youngman and Shin, Ik Jae and Tsukioka, Takuya",
    title   = "{Holographic QCD: Past, Present, and Future}",
    doi     = "10.1016/j.ppnp.2012.09.002",
    url     = "https://doi.org/10.1016/j.ppnp.2012.09.002",
    journal = "Prog. Part. Nucl. Phys.",
    volume  = "68",
    pages   = "55--112",
    year    = "2013"
}

@article{Adams:2012th,
    author  = "Adams, Allan and Carr, Lincoln D. and {Sch\"afer}, Thomas and Steinberg, Peter and Thomas, John E.",
    title   = "{Strongly Correlated Quantum Fluids: Ultracold Quantum Gases, Quantum Chromodynamic Plasmas, and Holographic Duality}",
    doi     = "10.1088/1367-2630/14/11/115009",
    url     = "https://doi.org/10.1088/1367-2630/14/11/115009",
    journal = "New J. Phys.",
    volume  = "14",
    pages   = "115009",
    year    = "2012"
}

@article{Vafa:2005ui,
    author        = "Vafa, Cumrun",
    title         = "{The String landscape and the swampland}",
    eprint        = "hep-th/0509212",
    archivePrefix = "arXiv",
    url           = "https://arxiv.org/abs/hep-th/0509212",
    year          = "2005"
}

@article{Ooguri:2006in,
    author        = "Ooguri, Hirosi and Vafa, Cumrun",
    title         = "{On the Geometry of the String Landscape and the Swampland}",
    eprint        = "hep-th/0605264",
    archivePrefix = "arXiv",
    doi           = "10.1016/j.nuclphysb.2006.10.033",
    url           = "https://doi.org/10.1016/j.nuclphysb.2006.10.033",
    journal       = "Nucl. Phys. B",
    volume        = "766",
    pages         = "21--33",
    year          = "2007"
}

@article{Arkani-Hamed:2006emk,
    author        = "Arkani-Hamed, Nima and Motl, Lubos and Nicolis, Alberto and Vafa, Cumrun",
    title         = "{The String landscape, black holes and gravity as the weakest force}",
    eprint        = "hep-th/0601001",
    archivePrefix = "arXiv",
    doi           = "10.1088/1126-6708/2007/06/060",
    url           = "https://doi.org/10.1088/1126-6708/2007/06/060",
    journal       = "JHEP",
    volume        = "06",
    pages         = "060",
    year          = "2007"
}

@article{Sadeghi:2021plz,
    author  = "Sadeghi, J. and Pourhassan, B. and Noori Gashti, S. and Upadhyay, S.",
    title   = "{Swampland conjecture and inflation model from brane perspective}",
    doi     = "10.1088/1402-4896/ac3a90",
    url     = "https://doi.org/10.1088/1402-4896/ac3a90",
    journal = "Phys. Scripta",
    volume  = "96",
    number  = "12",
    pages   = "125317",
    year    = "2021"
}

@article{Cordova:2022rer,
    author        = "Cordova, Clay and Ohmori, Kantaro and Rudelius, Tom",
    title         = "{Generalized Symmetry Breaking Scales and Weak Gravity Conjectures}",
    eprint        = "2202.05866",
    archivePrefix = "arXiv",
    primaryClass  = "hep-th",
    url           = "https://arxiv.org/abs/2202.05866",
    year          = "2022"
}

@article{Henriksson:2022oeu,
    author        = "Henriksson, Johan and McPeak, Brian and Russo, Francesco and Vichi, Alessandro",
    title         = "{Bounding Violations of the Weak Gravity Conjecture}",
    eprint        = "2203.09564",
    archivePrefix = "arXiv",
    primaryClass  = "hep-th",
    url           = "https://arxiv.org/abs/2203.09564",
    year          = "2022"
}

@article{Kaya:2022edp,
    author        = "Kaya, Saliha and Rudelius, Tom",
    title         = "{Higher-Group Symmetries and Weak Gravity Conjecture Mixing}",
    eprint        = "2202.04655",
    archivePrefix = "arXiv",
    primaryClass  = "hep-th",
    url           = "https://arxiv.org/abs/2202.04655",
    year          = "2022"
}

@article{Collazuol:2022jiy,
    author        = "Collazuol, Veronica and Gra\~na, Mariana and Herr\'aez, \'Alvaro",
    title         = "{E9 symmetry in the Heterotic String on $S^1$ and the Weak Gravity Conjecture}",
    eprint        = "2203.13354",
    archivePrefix = "arXiv",
    primaryClass  = "hep-th",
    url           = "https://arxiv.org/abs/2203.13354",
    year          = "2022"
}

@article{Capozziello:2011nr,
    author  = "Capozziello, Salvatore and De Laurentis, Mariafelicia and Odintsov, Sergei D. and Stabile, Arturo",
    title   = "{Hydrostatic equilibrium and stellar structure in f(R)-gravity}",
    doi     = "10.1103/PhysRevD.83.064004",
    url     = "https://doi.org/10.1103/PhysRevD.83.064004",
    journal = "Phys. Rev. D",
    volume  = "83",
    pages   = "064004",
    year    = "2011"
}

@article{McInnes:2022tut,
    author        = "McInnes, Brett",
    title         = "{Planar Black Holes as a Route to Understanding the Weak Gravity Conjecture}",
    eprint        = "2201.05257",
    archivePrefix = "arXiv",
    primaryClass  = "hep-th",
    url           = "https://arxiv.org/abs/2201.05257",
    year          = "2022"
}

@article{Cheung:2014vva,
    author        = "Cheung, Clifford and Remmen, Grant N.",
    title         = "{Naturalness and the Weak Gravity Conjecture}",
    eprint        = "1402.2287",
    archivePrefix = "arXiv",
    doi           = "10.1103/PhysRevLett.113.051601",
    url           = "https://doi.org/10.1103/PhysRevLett.113.051601",
    journal       = "Phys. Rev. Lett.",
    volume        = "113",
    pages         = "051601",
    year          = "2014"
}

@article{Rudelius:2022gyu,
    author        = "Rudelius, Tom",
    title         = "{Constraints on Early Dark Energy from the Axion Weak Gravity Conjecture}",
    eprint        = "2203.05581",
    archivePrefix = "arXiv",
    primaryClass  = "hep-th",
    url           = "https://arxiv.org/abs/2203.05581",
    year          = "2022"
}

@article{Cribiori:2022trc,
    author        = "Cribiori, Niccol\`o and Dall'Agata, Gianguido",
    title         = "{Weak gravity versus scale separation}",
    eprint        = "2203.05559",
    archivePrefix = "arXiv",
    primaryClass  = "hep-th",
    url           = "https://arxiv.org/abs/2203.05559",
    year          = "2022"
}

@article{Klaewer:2020lfg,
    author  = "Klaewer, Daniel and Lee, Seung-Joo and Weigand, Timo and Wiesner, Max",
    title   = "{Quantum corrections in 4d N = 1 infinite distance limits and the weak gravity conjecture}",
    doi     = "10.1007/JHEP03(2021)252",
    url     = "https://doi.org/10.1007/JHEP03(2021)252",
    journal = "JHEP",
    volume  = "03",
    pages   = "252",
    year    = "2021"
}

@article{Heidenreich:2015nta,
    author  = "Heidenreich, Ben and Reece, Matthew and Rudelius, Tom",
    title   = "{Sharpening the Weak Gravity Conjecture with Dimensional Reduction}",
    doi     = "10.1007/JHEP02(2016)140",
    url     = "https://doi.org/10.1007/JHEP02(2016)140",
    journal = "JHEP",
    volume  = "02",
    pages   = "140",
    year    = "2016"
}

@article{Heidenreich:2016aqi,
    author  = "Heidenreich, Ben and Reece, Matthew and Rudelius, Tom",
    title   = "{Evidence for a sublattice weak gravity conjecture}",
    doi     = "10.1007/JHEP08(2017)025",
    url     = "https://doi.org/10.1007/JHEP08(2017)025",
    journal = "JHEP",
    volume  = "08",
    pages   = "025",
    year    = "2017"
}

@article{Andriolo:2018lvp,
    author  = "Andriolo, Stefano and Junghans, Daniel and Noumi, Toshifumi and Shiu, Gary",
    title   = "{A Tower Weak Gravity Conjecture from Infrared Consistency}",
    doi     = "10.1002/prop.201800020",
    url     = "https://doi.org/10.1002/prop.201800020",
    journal = "Fortsch. Phys.",
    volume  = "66",
    number  = "5",
    pages   = "1800020",
    year    = "2018"
}

@article{Polchinski:2003bq,
    author        = "Polchinski, Joseph",
    title         = "{Monopoles, duality, and string theory}",
    eprint        = "hep-th/0304042",
    archivePrefix = "arXiv",
    doi           = "10.1142/S0217751X0401866X",
    url           = "https://doi.org/10.1142/S0217751X0401866X",
    journal       = "Int. J. Mod. Phys. A",
    volume        = "19S1",
    pages         = "145--156",
    year          = "2004"
}

@article{Ooguri:2016pdq,
    author  = "Ooguri, Hirosi and Vafa, Cumrun",
    title   = "{Non-supersymmetric AdS and the Swampland}",
    doi     = "10.4310/ATMP.2017.v21.n7.a8",
    url     = "https://doi.org/10.4310/ATMP.2017.v21.n7.a8",
    journal = "Adv. Theor. Math. Phys.",
    volume  = "21",
    pages   = "1787--1801",
    year    = "2017"
}

@article{Ibanez:2017kvh,
    author  = "Ibanez, Luis E. and Martin-Lozano, Victor and Valenzuela, Irene",
    title   = "{Constraining Neutrino Masses, the Cosmological Constant and BSM Physics from the Weak Gravity Conjecture}",
    doi     = "10.1007/JHEP11(2017)066",
    url     = "https://doi.org/10.1007/JHEP11(2017)066",
    journal = "JHEP",
    volume  = "11",
    pages   = "066",
    year    = "2017"
}

@article{Shiu:2016weq,
    author  = "Shiu, Gary and Soler, Pablo and Cottrell, William",
    title   = "{Weak Gravity Conjecture and extremal black holes}",
    doi     = "10.1007/s11433-019-9406-2",
    url     = "https://doi.org/10.1007/s11433-019-9406-2",
    journal = "Sci. China Phys. Mech. Astron.",
    volume  = "62",
    number  = "11",
    pages   = "110412",
    year    = "2019"
}

@article{Fisher:2017dbc,
    author        = "Fisher, Zachary and Mogni, Cynthia J.",
    title         = "{A Semiclassical, Entropic Proof of a Weak Gravity Conjecture}",
    eprint        = "1706.07957",
    archivePrefix = "arXiv",
    primaryClass  = "hep-th",
    url           = "https://arxiv.org/abs/1706.07957",
    year          = "2017"
}

@article{Cheung:2018cwt,
    author  = "Cheung, Clifford and Liu, Junyu and Remmen, Grant N.",
    title   = "{Proof of the Weak Gravity Conjecture from Black Hole Entropy}",
    doi     = "10.1007/JHEP10(2018)004",
    url     = "https://doi.org/10.1007/JHEP10(2018)004",
    journal = "JHEP",
    volume  = "10",
    pages   = "004",
    year    = "2018"
}

@article{Crisford:2017gsb,
    author  = "Crisford, Toby and Horowitz, Gary T. and Santos, Jorge E.",
    title   = "{Testing the Weak Gravity - Cosmic Censorship Connection}",
    doi     = "10.1103/PhysRevD.97.066005",
    url     = "https://doi.org/10.1103/PhysRevD.97.066005",
    journal = "Phys. Rev. D",
    volume  = "97",
    number  = "6",
    pages   = "066005",
    year    = "2018"
}

@article{Harlow:2015lma,
    author  = "Harlow, Daniel",
    title   = "{Wormholes, Emergent Gauge Fields, and the Weak Gravity Conjecture}",
    doi     = "10.1007/JHEP01(2016)122",
    url     = "https://doi.org/10.1007/JHEP01(2016)122",
    journal = "JHEP",
    volume  = "01",
    pages   = "122",
    year    = "2016"
}

@article{Hamada:2018dde,
    author  = "Hamada, Yuta and Noumi, Toshifumi and Shiu, Gary",
    title   = "{Weak Gravity Conjecture from Unitarity and Causality}",
    doi     = "10.1103/PhysRevLett.123.051601",
    url     = "https://doi.org/10.1103/PhysRevLett.123.051601",
    journal = "Phys. Rev. Lett.",
    volume  = "123",
    number  = "5",
    pages   = "051601",
    year    = "2019"
}

@article{Kinney:2018nny,
    author  = "Kinney, William H. and Vagnozzi, Sunny and Visinelli, Luca",
    title   = "{The zoo plot meets the swampland: mutual (in)consistency of single-field inflation, string conjectures, and cosmological data}",
    doi     = "10.1088/1361-6382/ab1d87",
    url     = "https://doi.org/10.1088/1361-6382/ab1d87",
    journal = "Class. Quant. Grav.",
    volume  = "36",
    number  = "11",
    pages   = "117001",
    year    = "2019"
}

@article{Harlow:2022gzl,
    author        = "Harlow, Daniel and Heidenreich, Ben and Reece, Matthew and Rudelius, Tom",
    title         = "{The Weak Gravity Conjecture: A Review}",
    eprint        = "2201.08387",
    archivePrefix = "arXiv",
    primaryClass  = "hep-th",
    url           = "https://arxiv.org/abs/2201.08387",
    year          = "2022"
}

@article{Saraswat:2016eaz,
    author  = "Saraswat, Prashant",
    title   = "{Weak gravity conjecture and effective field theory}",
    doi     = "10.1103/PhysRevD.95.025013",
    url     = "https://doi.org/10.1103/PhysRevD.95.025013",
    journal = "Phys. Rev. D",
    volume  = "95",
    number  = "2",
    pages   = "025013",
    year    = "2017"
}

@article{Nakayama:2015hga,
    author  = "Nakayama, Yu and Nomura, Yasunori",
    title   = "{Weak gravity conjecture in the AdS/CFT correspondence}",
    doi     = "10.1103/PhysRevD.92.126006",
    url     = "https://doi.org/10.1103/PhysRevD.92.126006",
    journal = "Phys. Rev. D",
    volume  = "92",
    number  = "12",
    pages   = "126006",
    year    = "2015"
}

@article{Bachlechner:2015qja,
    author  = "Bachlechner, Thomas C. and Long, Cody and McAllister, Liam",
    title   = "{Planckian Axions and the Weak Gravity Conjecture}",
    doi     = "10.1007/JHEP01(2016)091",
    url     = "https://doi.org/10.1007/JHEP01(2016)091",
    journal = "JHEP",
    volume  = "01",
    pages   = "091",
    year    = "2016"
}

@article{Bellazzini:2019xts,
    author  = "Bellazzini, Brando and Lewandowski, Matthew and Serra, Javi",
    title   = "{Positivity of Amplitudes, Weak Gravity Conjecture, and Modified Gravity}",
    doi     = "10.1103/PhysRevLett.123.251103",
    url     = "https://doi.org/10.1103/PhysRevLett.123.251103",
    journal = "Phys. Rev. Lett.",
    volume  = "123",
    number  = "25",
    pages   = "251103",
    year    = "2019"
}

@article{Aharony:2021mpc,
    author  = "Aharony, Ofer and Palti, Eran",
    title   = "{Convexity of charged operators in CFTs and the weak gravity conjecture}",
    doi     = "10.1103/PhysRevD.104.126005",
    url     = "https://doi.org/10.1103/PhysRevD.104.126005",
    journal = "Phys. Rev. D",
    volume  = "104",
    number  = "12",
    pages   = "126005",
    year    = "2021"
}

@article{Heidenreich:2019zkl,
    author  = "Heidenreich, Ben and Reece, Matthew and Rudelius, Tom",
    title   = "{Repulsive Forces and the Weak Gravity Conjecture}",
    doi     = "10.1007/JHEP10(2019)055",
    url     = "https://doi.org/10.1007/JHEP10(2019)055",
    journal = "JHEP",
    volume  = "10",
    pages   = "055",
    year    = "2019"
}

@article{Benakli:2020vng,
    author  = "Benakli, Karim and Branchina, Carlo and Lafforgue-Marmet, Gaetan",
    title   = "{U(1) mixing and the Weak Gravity Conjecture}",
    doi     = "10.1140/epjc/s10052-020-08691-4",
    url     = "https://doi.org/10.1140/epjc/s10052-020-08691-4",
    journal = "Eur. Phys. J. C",
    volume  = "80",
    number  = "12",
    pages   = "1118",
    year    = "2020"
}

@article{Lee:2018spm,
    author  = "Lee, Seung-Joo and Lerche, Wolfgang and Weigand, Timo",
    title   = "{A Stringy Test of the Scalar Weak Gravity Conjecture}",
    doi     = "10.1016/j.nuclphysb.2018.11.001",
    url     = "https://doi.org/10.1016/j.nuclphysb.2018.11.001",
    journal = "Nucl. Phys. B",
    volume  = "938",
    pages   = "321--350",
    year    = "2019"
}

@article{deAlwis:2019aud,
    author  = "de Alwis, S. and Eichhorn, A. and Held, A. and Pawlowski, J. M. and Schiffer, M. and Versteegen, F.",
    title   = "{Asymptotic safety, string theory and the weak gravity conjecture}",
    doi     = "10.1016/j.physletb.2019.134991",
    url     = "https://doi.org/10.1016/j.physletb.2019.134991",
    journal = "Phys. Lett. B",
    volume  = "798",
    pages   = "134991",
    year    = "2019"
}

@article{Andriolo:2020lul,
    author  = "Andriolo, Stefano and Huang, Tzu-Chen and Noumi, Toshifumi and Ooguri, Hirosi and Shiu, Gary",
    title   = "{Duality and axionic weak gravity}",
    doi     = "10.1103/PhysRevD.102.046008",
    url     = "https://doi.org/10.1103/PhysRevD.102.046008",
    journal = "Phys. Rev. D",
    volume  = "102",
    number  = "4",
    pages   = "046008",
    year    = "2020"
}

@article{Cremonini:2020smy,
    author  = "Cremonini, Sera and Jones, Callum R. T. and Liu, James T. and McPeak, Brian and Tang, Yuan",
    title   = "{NUT charge weak gravity conjecture from dimensional reduction}",
    doi     = "10.1103/PhysRevD.103.106011",
    url     = "https://doi.org/10.1103/PhysRevD.103.106011",
    journal = "Phys. Rev. D",
    volume  = "103",
    number  = "10",
    pages   = "106011",
    year    = "2021"
}

@article{Ibanez:2015fcv,
    author  = "Ibanez, Luis E. and Montero, Miguel and Uranga, Angel and Valenzuela, Irene",
    title   = "{Relaxion Monodromy and the Weak Gravity Conjecture}",
    doi     = "10.1007/JHEP04(2016)020",
    url     = "https://doi.org/10.1007/JHEP04(2016)020",
    journal = "JHEP",
    volume  = "04",
    pages   = "020",
    year    = "2016"
}

@article{Banerjee:2020xcn,
    author  = "Banerjee, Anirban and Cai, Hu and Heisenberg, Lavinia and Colg\'ain, Eoin O. and Sheikh-Jabbari, M. M. and Yang, Tao",
    title   = "{Hubble sinks in the low-redshift swampland}",
    doi     = "10.1103/PhysRevD.103.L081305",
    url     = "https://doi.org/10.1103/PhysRevD.103.L081305",
    journal = "Phys. Rev. D",
    volume  = "103",
    number  = "8",
    pages   = "L081305",
    year    = "2021"
}

@article{Kats:2006xp,
    author  = "Kats, Yevgeny and Motl, Lubos and Padi, Megha",
    title   = "{Higher-order corrections to mass-charge relation of extremal black holes}",
    doi     = "10.1088/1126-6708/2007/12/068",
    url     = "https://doi.org/10.1088/1126-6708/2007/12/068",
    journal = "JHEP",
    volume  = "12",
    pages   = "068",
    year    = "2007"
}

@article{a,
    author  = "Gashti, Saeed Noori and Pourhassan, Behnam and Sakall{\i}, \.{I}zzet",
    title   = "{Beyond extremality: Weak Gravity Conjecture constraints on gravitational lensing in gravity's rainbow}",
    doi     = "10.1007/JHEP04(2026)134",
    url     = "https://doi.org/10.1007/JHEP04(2026)134",
    journal = "JHEP",
    volume  = "2026",
    number  = "4",
    pages   = "134",
    year    = "2026"
}

@article{c,
    author        = "Noori Gashti, Saeed and others",
    title         = "{Testing the Weak Gravity Conjecture via Gravitational Lensing, Black Hole Shadows, and Barrow Thermodynamics in F(R)-Euler-Heisenberg (A)dS Black Holes}",
    eprint        = "2602.XXXXX",
    archivePrefix = "arXiv",
    primaryClass  = "gr-qc",
    year          = "2026",
    note          = "arXiv e-prints"
}

@book{d,
    author    = "Noori Gashti, Saeed and Pourhassan, Behnam",
    title     = "{Swampland Program: Cosmic-Quantum Connection}",
    publisher = "Damghan University Press",
    year      = "2026"
}

@article{f,
    author  = "Gashti, Saeed Noori and Afshar, Mohammad Ali S. and Alipour, Mohammad Reza and Sakall{\i}, \.{I}zzet and Pourhassan, Behnam and Sadeghi, Jafar",
    title   = "{Weak gravity conjecture in ModMax black holes: weak cosmic censorship and photon sphere analysis}",
    doi     = "10.1140/epjc/s10052-025-14890-8",
    url     = "https://doi.org/10.1140/epjc/s10052-025-14890-8",
    journal = "Eur. Phys. J. C",
    volume  = "85",
    number  = "10",
    pages   = "1144",
    year    = "2025"
}

@article{i,
    author        = "Alipour, Mohammad Reza and others",
    title         = "{Reconciling the weak gravity and weak cosmic censorship conjectures in Einstein-Euler-Heisenberg-AdS black holes}",
    eprint        = "2504.03453",
    archivePrefix = "arXiv",
    primaryClass  = "gr-qc",
    url           = "https://arxiv.org/abs/2504.03453",
    year          = "2025"
}

@article{Urbano:2018kax,
    author        = "Urbano, Alfredo",
    title         = "{Towards a proof of the Weak Gravity Conjecture}",
    eprint        = "1810.11001",
    archivePrefix = "arXiv",
    primaryClass  = "hep-th",
    url           = "https://arxiv.org/abs/1810.11001",
    year          = "2018"
}

@article{Hod:2017uqc,
    author  = "Hod, Shahar",
    title   = "{A proof of the weak gravity conjecture}",
    doi     = "10.1142/S0218271817420044",
    url     = "https://doi.org/10.1142/S0218271817420044",
    journal = "Int. J. Mod. Phys. D",
    volume  = "26",
    number  = "12",
    pages   = "1742004",
    year    = "2017"
}

@article{Hod:2006jw,
    author  = "Hod, Shahar",
    title   = "{Universal Bound on Dynamical Relaxation Times and Black-Hole Quasinormal Ringing}",
    doi     = "10.1103/PhysRevD.75.064013",
    url     = "https://doi.org/10.1103/PhysRevD.75.064013",
    journal = "Phys. Rev. D",
    volume  = "75",
    pages   = "064013",
    year    = "2007"
}

@article{Hod:2010hw,
    author  = "Hod, Shahar",
    title   = "{Relaxation dynamics of charged gravitational collapse}",
    doi     = "10.1016/j.physleta.2010.05.024",
    url     = "https://doi.org/10.1016/j.physleta.2010.05.024",
    journal = "Phys. Lett. A",
    volume  = "374",
    pages   = "2901",
    year    = "2010"
}

@article{1,
    author        = "Yerra, Pavan Kumar and Bhamidipati, Chandrasekhar",
    title         = "{Topological charge and black hole photon spheres in massive gravity}",
    eprint        = "2509.04261",
    archivePrefix = "arXiv",
    primaryClass  = "gr-qc",
    url           = "https://arxiv.org/abs/2509.04261",
    year          = "2025"
}

@article{2,
    author = "Sucu, Erdem and Dengiz, Suat and Sakall{\i}, \.{I}zzet",
    title  = "{Thermal and Optical Signatures of Einstein-Dyonic ModMax Black Holes with GUP and Plasma Modifications}",
    note   = "SSRN preprint 5894291",
    url    = "https://ssrn.com/abstract=5894291",
    year   = "2025"
}

@article{WOS:001248758300001,
    author  = "Sadeghi, J. and Noori Gashti, S. and Sakall{\i}, \.{I}. and Pourhassan, B.",
    title   = "{Weak gravity conjecture of charged-rotating-AdS black hole surrounded by quintessence and string cloud}",
    doi     = "10.1016/j.nuclphysb.2024.116581",
    url     = "https://doi.org/10.1016/j.nuclphysb.2024.116581",
    journal = "Nucl. Phys. B",
    volume  = "1004",
    pages   = "116581",
    year    = "2024"
}

@article{WOS:001062453500003,
    author  = "Uniyal, Akhil and Kanzi, Sara and Sakall{\i}, \.{I}zzet",
    title   = "{Some observable physical properties of the higher dimensional dS/AdS black holes in Einstein-bumblebee gravity theory}",
    doi     = "10.1140/epjc/s10052-023-11846-8",
    url     = "https://doi.org/10.1140/epjc/s10052-023-11846-8",
    journal = "Eur. Phys. J. C",
    volume  = "83",
    number  = "7",
    pages   = "668",
    year    = "2023"
}

@article{WOS:001061803400004,
    author  = "Sadeghi, J. and Pourhassan, B. and Noori Gashti, S. and Sakall{\i}, \.{I}. and Alipour, M. R.",
    title   = "{de Sitter swampland conjecture in string field inflation}",
    doi     = "10.1140/epjc/s10052-023-11822-2",
    url     = "https://doi.org/10.1140/epjc/s10052-023-11822-2",
    journal = "Eur. Phys. J. C",
    volume  = "83",
    number  = "7",
    pages   = "635",
    year    = "2023"
}

@article{WOS:000387373600001,
    author  = "Montero, Miguel and Shiu, Gary and Soler, Pablo",
    title   = "{The Weak Gravity Conjecture in three dimensions}",
    doi     = "10.1007/JHEP10(2016)159",
    url     = "https://doi.org/10.1007/JHEP10(2016)159",
    journal = "JHEP",
    volume  = "10",
    pages   = "159",
    year    = "2016"
}

@article{WOS:000380409200053,
    author  = "Kooner, Karta and Parameswaran, Susha and Zavala, Ivonne",
    title   = "{Warping the Weak Gravity Conjecture}",
    doi     = "10.1016/j.physletb.2016.05.082",
    url     = "https://doi.org/10.1016/j.physletb.2016.05.082",
    journal = "Phys. Lett. B",
    volume  = "759",
    pages   = "402--409",
    year    = "2016"
}

@article{WOS:000853339300007,
    author  = "Palti, Eran and Sharon, Adar",
    title   = "{Convexity of charged operators in CFTs with multiple Abelian symmetries}",
    doi     = "10.1007/JHEP09(2022)078",
    url     = "https://doi.org/10.1007/JHEP09(2022)078",
    journal = "JHEP",
    volume  = "09",
    pages   = "078",
    year    = "2022"
}

@article{WOS:000487936500013,
    author  = "Bonnefoy, Quentin and Dudas, Emilian and {L\"ust}, Severin",
    title   = "{On the weak gravity conjecture in string theory with broken supersymmetry}",
    doi     = "10.1016/j.nuclphysb.2019.114738",
    url     = "https://doi.org/10.1016/j.nuclphysb.2019.114738",
    journal = "Nucl. Phys. B",
    volume  = "947",
    pages   = "114738",
    year    = "2019"
}

@article{WOS:000904391700001,
    author  = "Sadeghi, Jafar and Alipour, Mohammad Reza and Noori Gashti, Saeed",
    title   = "{Scalar Weak Gravity Conjecture in Super Yang-Mills Inflationary Model}",
    doi     = "10.3390/universe8120621",
    url     = "https://doi.org/10.3390/universe8120621",
    journal = "Universe",
    volume  = "8",
    number  = "12",
    pages   = "621",
    year    = "2022"
}

@article{WOS:001620407300001,
    author  = "Gashti, Saeed Noori and Afshar, Mohammad Ali S. and Alipour, Mohammad Reza and Pourhassan, Behnam and Sadeghi, Jafar",
    title   = "{From minimal Higgs inflation with $R^2$ term in Palatini gravity to Swampland conjectures under ACT constraints}",
    doi     = "10.1140/epjc/s10052-025-15066-0",
    url     = "https://doi.org/10.1140/epjc/s10052-025-15066-0",
    journal = "Eur. Phys. J. C",
    volume  = "85",
    number  = "11",
    pages   = "1343",
    year    = "2025"
}

\end{document}